\documentclass[
reprint,
 amsmath,amssymb,
 aps,
prl,
floatfix
]{revtex4-2}

\usepackage{graphicx}
\usepackage{dcolumn}
\usepackage{bm}
\usepackage{aas_macros}
\usepackage{xcolor}
\usepackage{mathtools}

\definecolor{green}{rgb}{0.05,0.6,0}
\usepackage{hyperref}
\hypersetup{colorlinks=true,linkcolor=red,citecolor=blue,urlcolor=cyan}

\begin{document}

\title{Dynamical tides in neutron stars with first-order phase transitions: the role of the discontinuity mode} 

\author{Jonas P. Pereira$^{1,2,3,4}$, Lucas Tonetto$^{5}$, Micha{\l} Bejger$^{3,6}$, J. Leszek Zdunik$^{3}$, Pawe{\l} Haensel$^{3}$}
\email{e-mails to: jonas.pereira@unb.br, bejger@camk.edu.pl}
\affiliation{$^{1}$ Institute of Physics \& International Center of Physics, University of Brasilia, 70297-400, Brasilia, Federal District, Brazil}
\affiliation{$^{2}$ Departamento de Astronomia, Instituto de Astronomia, Geof\' isica e Ci\^encias Atmosf\'ericas (IAG), Universidade de S\~ao Paulo, S\~ao Paulo, 05508-090, Brazil}
\affiliation{$^{3}$ Nicolaus Copernicus Astronomical Center, Polish Academy of Sciences, Bartycka 18, 00-716, Warsaw, Poland}
\affiliation{$^{4}$ Programa de Pós-Graduação em Astrofísica, Cosmologia e Gravitação (PPGCosmo), Federal University of Espírito Santo, Vitória-ES, 29075-910, Brazil}
\affiliation{$^{5}$ Dipartimento di Fisica, ``Sapienza'' Universit\`a di Roma \& Sezione INFN Roma1, P.A. Moro 5, 00185, Roma, Italy}
\affiliation{$^{6}$ INFN Sezione di Ferrara, Via Saragat 1, 44122 Ferrara, Italy}

\date{\today}

\begin{abstract}
\noindent During the late stages of a binary neutron star inspiral, dynamical tides induced in each star by its companion become significant and should be included in complete gravitational-wave (GW) modeling. We investigate the coupling between the tidal field and quasi-normal modes in hybrid stars and show that the discontinuity mode ($g$-mode)—intrinsically associated with first-order phase transitions and buoyancy—contributes non-negligibly compared with the fundamental $f$-mode. We find that the $g$-mode overlap integral can reach up to $\sim 10\%$ of the $f$-mode value for hybrid star masses in the range 1.4–2.0$M_{\odot}$, with the largest values generally associated with larger density jumps. This leads to a GW phase shift due to the $g$-mode of $\Delta \phi_g \lesssim 0.1$–$1$ rad (i.e., up to $\sim$5\%–10\% of $\Delta \phi_f$), with the largest shifts occurring for masses near the phase transition. At higher masses, the shifts remain smaller and nearly constant, with $\Delta \phi_g \lesssim 0.1$ rad (roughly $\sim 1\%$ of $\Delta \phi_f$). These GW shifts may be relevant even at the design sensitivity of current second-generation GW detectors in the most optimistic cases. Moreover, if a $g$-mode is present and lies near the $f$-mode frequency, neglecting it in the GW modeling can lead to systematic biases in neutron star parameter estimation, resulting in radius errors of up to $1\%–2\%$. These results show the importance of dynamical tides to probe neutron stars' equation of state, and to test the existence of dense-matter phase transitions.
\end{abstract}

\maketitle

\noindent {\bf {\it Introduction.---}}Important constraints on the structure of neutron stars (NSs) are emerging from electromagnetic (EM) and gravitational wave (GW) observations. Concerning GWs, the binary NS event GW170817 \citep{2017PhRvL.119p1101A,2019PhRvX...9a1001A} has already provided important limits on NS tidal deformability: a softer equation of state (EOS) at intermediate densities is preferred; however, many EOS candidates remain due to measurement uncertainties. Current second generation (2G) GW detectors, the Advanced LIGO \cite{2015CQGra..32g4001L}, Advanced Virgo \cite{2015CQGra..32b4001A}, and KAGRA \cite{2019NatAs...3...35K}, 2.5G detectors like the LIGO Voyager \citep{2020CQGra..37p5003A} and NEMO \citep{2020PASA...37...47A}, and 3G detectors, such as the Cosmic Explorer (CE) \cite{2019BAAS...51g..35R} and the Einstein Telescope (ET) \cite{2020JCAP...03..050M,2025arXiv250312263A} promise to significantly reduce these uncertainties. The inferred NS radius uncertainties will be around a few hundred meters (few percent in relative terms) for several detections with 3G detectors (see, e.g., \citep{2021arXiv210812368C,2024PhRvD.110d3013W,2025arXiv250312263A} and references therein), needing {\it perforce} more detailed models to describe the data.

Significant progress has also been made on the EM side using light curve analysis and ray-tracing techniques. NICER \cite{NICER}, currently observing several NSs, can already provide radius measurements with 10$\%$ uncertainties at 68$\%$ probability \citep{2021ApJ...918L..28M}. Combining NICER measurements results in even smaller radius uncertainties, reaching around 5$\%$ when GW constraints are also considered \citep{2021ApJ...918L..28M,2024ApJ...971L..19R}. Near-future missions such as STROBE-X \citep{2019arXiv190303035R}, ATHENA \cite{2013arXiv1306.2307N}, and eXTP \cite{2016SPIE.9905E..1QZ,2019SCPMA..6229503W}, promise even smaller NS radius uncertainties, around a few percent. NICER’s modeling of the EM emission from hot spots on NS surfaces suggests that stiffer EOSs are preferred, as stars with 1.35 $M_{\odot}$ and 2.0 $M_{\odot}$ have similar radii \citep{2019ApJ...887L..21R,2021ApJ...918L..28M,2021arXiv210506980R,2019ApJ...887L..24M}. So far, these measurements are consistent with GW observations \citep{2021ApJ...918L..28M,2021arXiv210506981R}. However, if future experiments confirm the PREX-II results \citep{2021PhRvL.126q2502A,2021PhRvL.126q2503R}, tension may arise \citep{2021PhRvL.126q2503R,2021arXiv210813071J}.
It could be resolved by dense-matter phase transitions \citep{2021PhRvL.126q2503R,2021arXiv210813071J}---theoretically allowed by Quantum Chromodynamics (QCD) \citep{2008RvMP...80.1455A}---leading to the concept of hybrid stars (see, e.g., \citep{2018ApJ...860...12P,2023PhRvD.107j3042R} and references therein). 

The existence of hybrid stars could also lead to phenomena that falsify some models when measurements become more precise, such as unique fingerprints on the GW waveforms of postmerger NS binaries \citep{2019PhRvL.122f1102B,2020PhRvL.124q1103W}, impacts on static tidal deformations and NS radii \citep{2018PhRvL.120z1103M,2023PhRvD.108b3010R,2023PhRvL.130t1403R}, and characteristic bursts of EM energy \citep{2003ApJ...586.1250B,2007A&A...465..533Z,2008A&A...479..515Z}. 

More sensitive GW detectors will fully cover the binary inspiral, merger and postmerger \citep{2020GReGr..52..109C}, the latter currently limited by the lack of sensitivity at GW frequencies above ${\sim}\,1$ kHz. In particular, for the Advanced LIGO and the Advanced Virgo detectors at their design sensitivity, it will be possible to measure tidal aspects of the late inspiral, where the adiabatic approximation is no longer accurate and dynamical tides \citep{1994MNRAS.270..611L,2020PhRvD.101f3007E,2021MNRAS.504.1273P,2019PhRvD.100b1501S,2021MNRAS.503..533A} become relevant \citep{2022PhRvL.129h1102P,2022PhRvD.105l3032W}. 
The main implication of dynamical tides is that stellar modes could resonate, extracting energy from the orbit and altering the GW waveform \citep{1994MNRAS.270..611L}. For NSs without phase transitions, the mode that most contributes to dynamical tides is the fundamental one \citep{1994MNRAS.270..611L}. However, when other phenomena are concerned, other modes could also be relevant \citep{2012PhRvL.108a1102T,2020PhRvD.101h3002S,2021MNRAS.508.1732K,2021MNRAS.504.1273P,2022MNRAS.513.4045K,2022MNRAS.514.1628K}. 

Dynamical tides in hybrid stars have only recently begun to be investigated for the discontinuity $g$-modes \citep{finn1986,2001PhRvD..65b4010S,2003MNRAS.338..389M,2020PhRvD.101l3029T}, also called interface mode ($i$-mode), and their potential detectability has already been stressed \citep{2021PhRvD.103f3015L,2024ApJ...964...31M,2025arXiv250406181C}. Our goal is to investigate discontinuity $g$-modes in a more systematic and detailed way. To do so, we use EOSs that span a broad range of density jumps and satisfy all astrophysical constraints. We also apply highly precise methods to compute the eigenmodes. Finally, we explicitly account for all boundary conditions on the perturbations—which are essential for the self-consistency of the problem and reliable estimates—and carry out our analysis within full general relativity (similarly to \citep{2024ApJ...964...31M} and differently from \citep{2025arXiv250406181C}). This allows us to assess the conditions under which $g$-modes could compete with the fundamental $f$-mode—and to understand the consequences of neglecting them.

We show here that the discontinuity $g$-mode, intrinsically associated with a first-order phase transition and buoyancy, can contribute non-negligibly to GW phase shifts when compared to the fundamental $f$-mode, and it might even be detected by current GW detectors for larger frequencies if they reach their design sensitivity. The largest shifts [$\lesssim (0.1-1)$ rad] happen for stellar masses close to the phase transition mass---a novelty of our analysis---, and almost constant, moderate shifts ($\lesssim 0.1$ rad) occur for larger NS masses. This differs from \cite{2024ApJ...964...31M}, which suggests detectability with aLIGO for smaller frequencies \footnote{We note that \cite{2024ApJ...964...31M}'s modeling approach is broadly similar in spirit to ours: they also treat the density jump as a free parameter and adopt a constant-speed-of-sound model for the quark phase ($c_s^2=1$). However, there are significant differences in the choice of hadronic equations of state (EOSs). While they employ either a very stiff or soft hadronic model, our hadronic EOS lies within a more moderate and observationally supported range, consistent with tidal deformability constraints. Thus, a direct comparison of our results with theirs is not straightforward. In addition, the authors do not provide details of their numerical integration scheme.}, and from \citep{2025arXiv250406181C}, which show on their Fig. 2 (in the main text and in the Supplemental Material) that larger density jumps lead to greater changes in the GW phase shift, and that the phase shift generally decreases with the NS mass increase. Our Figs.~\ref{fig:overlap_integrals} and \ref{fig:Number_cycles_variation} imply a more complicated relation between the phase shift, density jump and NS mass. In particular---and as a second novelty---near the transition mass small density jumps produce nonlinear, non-negligible GW phase shifts. Finally, we show that neglecting the $g$-mode GW phase shift can bias stellar-parameter inference, with important consequences for high-precision facilities such as eXTP and the Einstein Telescope.

\noindent{\bf {\it Results.---}}The EOSs we use provide the $M$-$R$ relation illustrated in Fig.~\ref{fig: MR_relation}. Details about them are given in the Supplemental Material. Our selected EOSs have phase transitions occurring at a mass just below $1.4\,M_{\odot}$ ($\sim 1.33M_{\odot}$). All EOSs exhibit a baryon number density jump $n_q/n_h>1$ at the quark-hadron interface, equivalently shown in our plots in terms of $n_q/n_h-1>0$. The EOSs share a common transition density at the base of the hadronic phase [$\epsilon_h \simeq 2.2\,\epsilon_{\rm{sat}}$ ($n_h \simeq 2.1\,n_{\rm{sat}}$), with $\epsilon_{\rm{sat}} = 2.7 \times 10^{14}$ g cm$^{-3}$ and $n_{\rm{sat}}=0.16$ fm$^{-3}$], and feature a corresponding phase transition pressure of $p_t = 6.71 \times 10^{34}$ dyn cm$^{-2}$ (which is in agreement with $p(2\epsilon_{\rm{sat}})=3.5^{+2.7}_{-1.7}\times 10^{34}$\ dyn\,cm$^{-2}$, coming from GW170817 \citep{2018PhRvL.121p1101A}); they are also consistent with astrophysical constraints in multiple contexts \citep{2021PhRvC.103c5802X,2021ApJ...913...27L,2025ApJ...983...17H,2020NatPh..16..907A,2024PhRvD.109f3035C,2025PhRvD.111f3007M}. To capture a wide range of scenarios, we explore both small and large jumps ($1.1 \leq n_q/n_h \leq 1.9$), encompassing weak and strong phase transitions \cite{2008A&A...479..515Z,2019A&A...622A.174S}. We adopt representative mass values of $1.4\,M_{\odot}$, $1.8\,M_{\odot}$, and $2.0\,M_{\odot}$ in our analysis, as indicated by the horizontal dashed lines in Fig.~\ref{fig: MR_relation}. The extent of the quark core slowly increases with stellar mass across the full range of $n_q/n_h$ considered, $1.1$–$1.9$. For $1.4\,M_{\odot}$ stars, it ranges from approximately $25\%$ to $55\%$ of the stellar radius; for $1.8\,M_{\odot}$, from $55\%$ to $75\%$; and for $2.0\,M_{\odot}$, from $60\%$ to $80\%$; see the Supplemental Material for more details.

\begin{figure}
\includegraphics[width=\columnwidth]{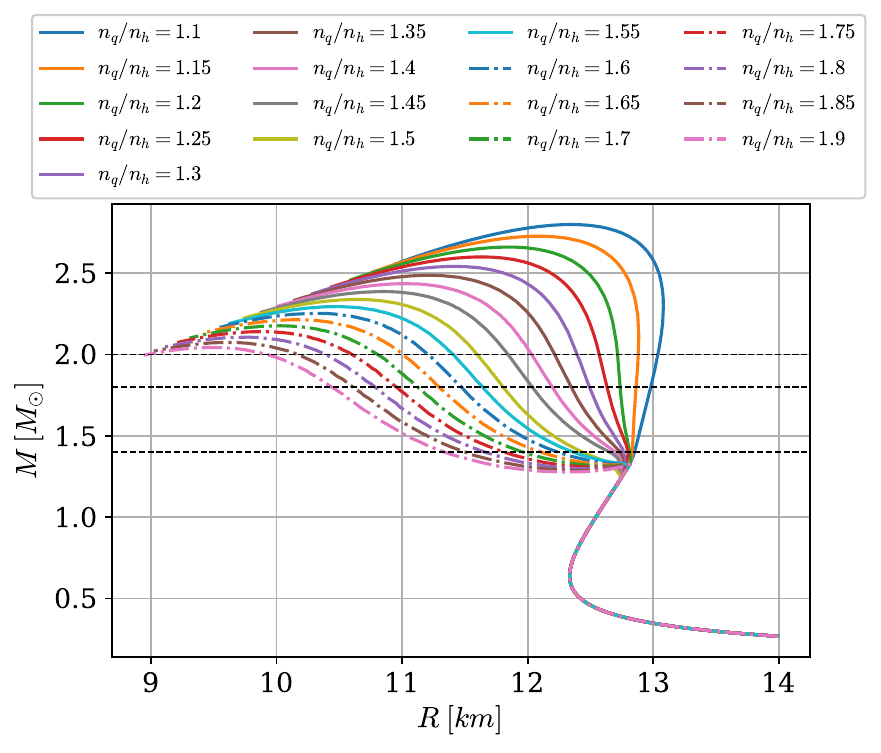}
\caption{Mass-radius relations for hybrid stars for various EOSs characterized by $n_q/n_h$ values ranging from 1.1 to 1.9, spanning scenarios from weak to strong phase transitions.
To facilitate identification of the properties of the stellar models analyzed, dashed lines have been plotted at $1.4\,M_{\odot}$, $1.8\,M_{\odot}$, and $2.0\,M_{\odot}$.
}
\label{fig: MR_relation}
\end{figure}

We make now some estimates for dynamical tides. We follow a ``hybrid'' approach where we take as a basis some Newtonian expressions but consider GR corrections to their terms. This strategy is motivated by the fact that extending the spectral decomposition method---commonly used in the Newtonian treatment of dynamical tides---to full GR remains a highly nontrivial task \cite{2024PhRvD.109f4004P}. When energy is extracted from the tidal field into the modes of a star, it affects the number of orbital cycles (the GW phase) before the coalescence \citep{1994MNRAS.270..611L}:
\begin{flalign}
\begin{aligned}
\label{N_change_main}
    \Delta N_{nl}&=\frac{\Delta \phi_{nl}}{4\pi}\simeq - 4.8 \times 10^{-3}\left( \frac{\nu_n}{10^{3}\,\rm{Hz}}\right)^{-2} \times \\
    &\times \left( \frac{Q_{nl}}{10^{-2}}\right)^{2} M_{1.4}^{-4}R_{10}^2 \left(\frac{2q}{1+q} \right),
\end{aligned}
\end{flalign}
where the quantity $n$ is associated with various modes in the star, for instance $f$-, $g$-, $s$- and $p$-modes, 
$\nu_n$ is the frequency of these modes, $q\equiv M'/M$ is the mass ratio of the binary, $M_{1.4}\equiv M/(1.4\,M_\odot)$ and $R_{10}\equiv R/(10\,\mathrm{km})$. For definiteness, we fix the companion NS mass to $M' = 1.4M_{\odot}$.
In addition, $Q_{nl}$ is the dimensionless overlap integral, which quantifies the strength of the coupling between the tidal field and the stellar oscillation mode $\nu_n$. Finally, $\Delta \phi_{nl}$ is the GW phase shift due to the resonance of a mode with frequency $\nu_n$ and overlap integral $Q_{nl}$. For $Q_{nl}$, we adopt the relativistic generalization of the Newtonian overlap integral \cite{2021MNRAS.506.2985K}: 
\begin{flalign}
\begin{aligned}
    Q_{nl} 
    &\equiv \frac{1}{MR^l} \int d^3x  \sqrt{-g} e^{-\nu} (\epsilon + p) \bar{\xi}^i_n \nabla_i (r^l Y_{m}^l) \\
    &= \frac{1}{MR^l}\int_0^R dr\, e^{\lambda} \, (\epsilon + p)\, r^{2l} \left[ l e^{-\lambda} W_{nl} - l(l+1) V_{nl} \right], 
    \label{overlap_integral}
\end{aligned}
\end{flalign}
where $g$ is the determinant of the background metric, $\epsilon(p)$ is the energy density(pressure), and $\xi^i_n\equiv \left(r^{l-1} e^{-\lambda}W_{nl}(r) Y_{m}^l, -r^{l-2} V_{nl}(r) \partial_\theta Y^l_m, -\frac{r^{l-2}}{\sin^2\theta} V_{nl}(r) \partial_\phi Y^l_m\right)$. The overbar in $\bar{\xi}^i_n$ denotes the complex conjugate of $\xi^i_n$. Each mode $\xi^i_n$ is normalized according to $MR^2=\int d^3x\sqrt{-g} e^{-\nu} (\epsilon + p) \bar{\xi}^i_n \xi_{ni}=\int_0^R dr\, e^{\lambda} \, (\epsilon + p)\, r^{2l} \left[W_{nl}^2 + l(l+1) V_{nl}^2 \right]$.
Further details about the quasi-normal modes in GR are provided in the Supplemental Material.

From now on, we assume $l=2$, which corresponds to the dominant multipole in dynamical tides. Figure~\ref{fig:overlap_integrals} shows the overlap integrals of the $f$- and $g$-modes for our EOS models and NSs with masses of $1.4\,M_{\odot}$, $1.8\,M_{\odot}$, and $2.0\,M_{\odot}$; to simplify the notation, we omit the $l=2$ index in $Q_{nl}$ and related quantities. It is clear a nontrivial feature of $Q_g$ for masses close to the transition mass ($\sim 1.4\,M_{\odot}$): it can take negative and null values. This behavior stems from the distinct structure of the $g$-mode eigenfunctions $V$ and $W$, which differ significantly from those of the $f$-mode and could even change sign, and the small size of the quark phase. Further details are given in the Discussion and the Supplemental Material. The corresponding $\Delta \phi_{n}$ values are shown in Fig.~\ref{fig:Number_cycles_variation}. We find in general $\Delta \phi_{g} \lesssim (0.1–1)$ rad, with the largest shifts for masses near the phase transition one and strong phase transitions. Note also that for very small density jumps, $\Delta \phi_g \sim (1–10)$ rad, resulting from the large negative values of $Q_g$ shown in Fig.~\ref{fig:overlap_integrals}.

\begin{figure}
\includegraphics[width=\columnwidth]{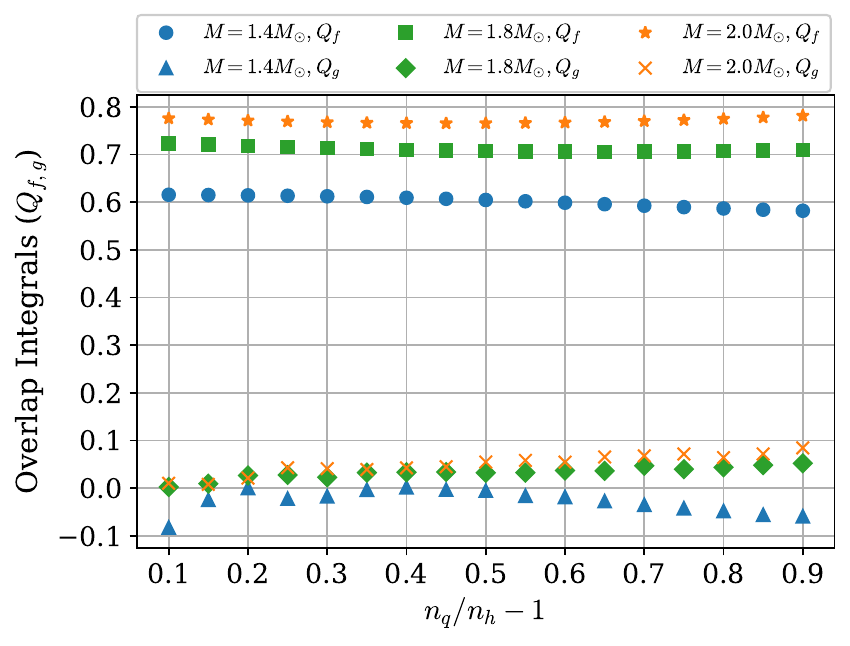}
\caption{Overlap integrals for the $f$- and $g$-modes for $1.4\,M_{\odot}$, $1.8\,M_{\odot}$ and $2.0\,M_{\odot}$ hybrid stars. As expected, $Q_f$ is dominant (due to the mode's influence on the entire star and its lack of nodes), but $Q_g$ is only approximately a factor of 10 smaller. The variations for $1.4\,M_{\odot}$ case are a consequence of the small quark phase and how it nonlinearly influences the eigenfrequencies, even allowing them to change sign in the hadronic phase (see the Supplemental Material). }
\label{fig:overlap_integrals}
\end{figure}

\begin{figure}
\includegraphics[width=\columnwidth]{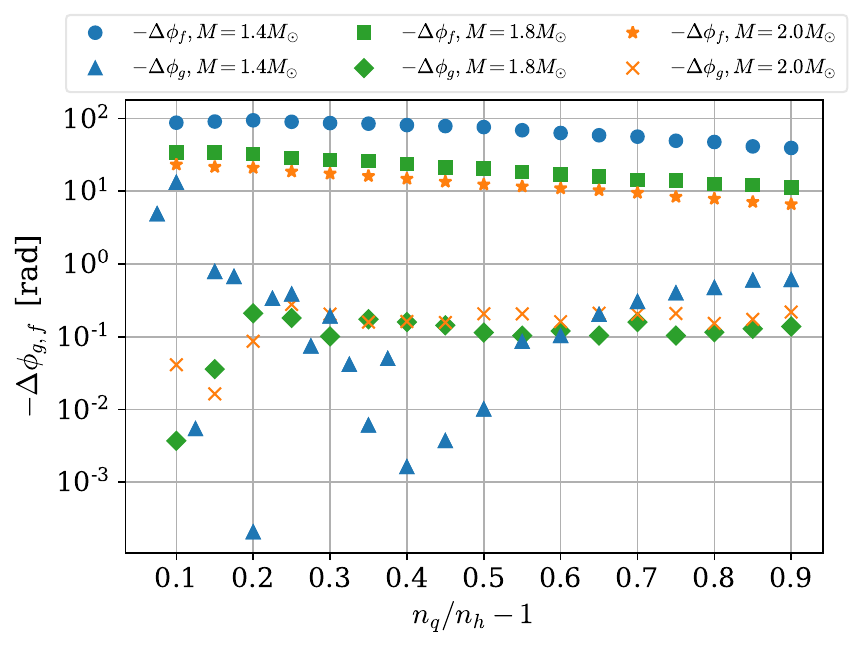}
\caption{GW phase shift for the $f$- and $g$-modes for a stars with $1.4\,M_{\odot}$, $1.8\,M_{\odot}$ and $2.0\,M_{\odot}$ and different number baryon density jumps. We set the NS companion mass to $1.4\,M_{\odot}$. We adopted a denser sampling in $n_q/n_h-1$ (step $0.025$) for the $1.4M_\odot$ sequence over the range $n_q/n_h - 1 \leq 0.4$ because this mass lies closest to the phase-transition threshold in our models.
}
\label{fig:Number_cycles_variation}
\end{figure}

\begin{figure}
\includegraphics[width=\columnwidth]{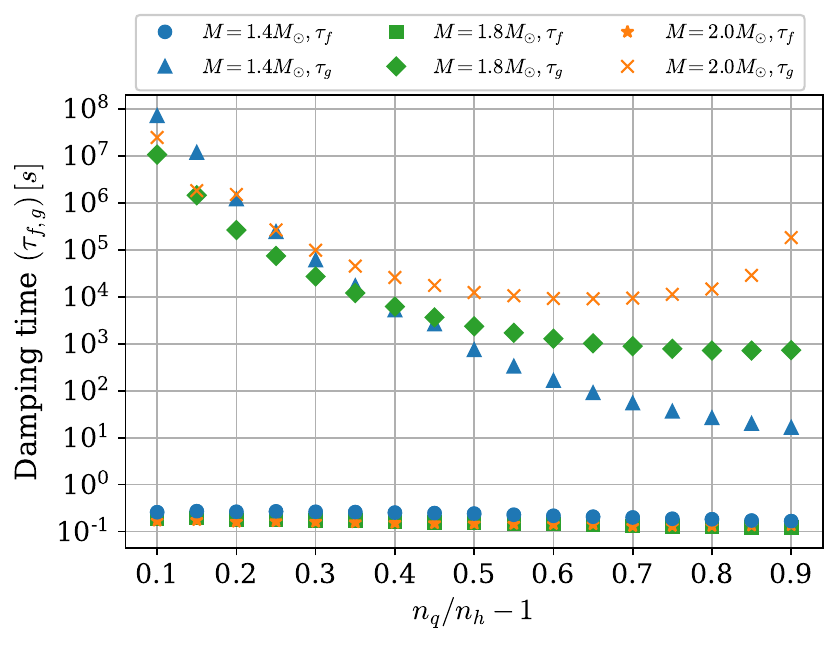}
\caption{Damping times for the $f$- and $g$-modes of hybrid stars with masses of $1.4\,M_{\odot}$, $1.8\,M_{\odot}$, and $2.0\,M_{\odot}$. The damping times of $g$-modes display nonlinear behavior with $n_q/n_h$, owing to their dependence on buoyancy. These damping times span roughly seven orders of magnitude and can reach values as large as $\sim (10–100)s$ for hybrid stars with significant density jumps.
}
\label{damping}
\end{figure}

The damping times for the $f$- and $g$-modes associated with the EOSs with varying density jumps are shown in Fig.~\ref{damping}. These times are key to understanding mode persistence in stars. It is well known that $f$-modes have damping times around $10^{-1}s$, whereas $g$-modes are highly sensitive to the density jump and transition pressure \citep{PhysRevD.98.124014,mariani2022}, decreasing nonlinearly with increasing jump. The stability of the $f$-mode stems from its dependence on compactness \citep{1998MNRAS.299.1059A}, which changes little among stars of the same mass but different $n_q/n_h$. For the models and masses considered, $g$-mode damping times range from $10^8s$ to $\sim 10^1s$, indicating that stellar $g$-mode deformations may persist up to merger, potentially enabling the conversion of vibrational energy into other forms.

\begin{figure}
\includegraphics[width=\columnwidth]{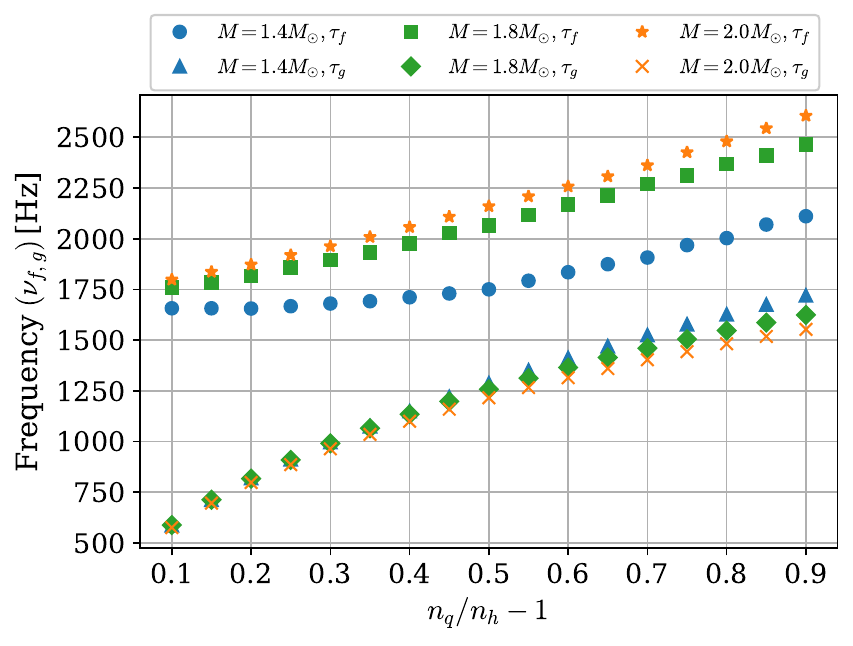}
\caption{Frequencies for the $f$-mode and $g$-mode for stars with $1.4\,M_{\odot}$, $1.8\,M_{\odot}$ and $2.0\,M_{\odot}$  and different number baryon density jumps.}
\label{jump_frequency}
\end{figure}

We now consider the case of two resonant modes with closely spaced frequencies, such that their individual contribution to the total number of orbital cycles due to a star, $\Delta N$, and the associated GW phase shift, $\Delta \phi$, cannot be individually resolved. This situation may arise if at least one of the modes has a long damping time, allowing it to remain active over a larger frequency range.  In this case, $\Delta N\approx  \sum_n \Delta N_{n}$ and $\Delta \phi \approx \sum_n \Delta \phi_{n}$. We will focus on $n=f,g$ and large density jumps, where this might happen. For simplicity, we take $\nu_g=a_g\nu_f$, $Q_{g}=q_gQ_{f}$ and $(a_g,q_g)<1$. Thus, from Eq.~\eqref{N_change_main}, taking into account the contributions of both $f$- and $g$-modes, $\Delta N$ is:
\begin{flalign}
\begin{aligned}
    \Delta N &= \frac{\Delta \phi}{4\pi}\propto \left(\frac{Q_{f}}{\nu_f}\right)^2 
    + \left(\frac{Q_{g}}{\nu_g}\right)^2\equiv \left(\frac{Q_{f}}{\tilde{\nu}_f}\right)^2, \\ &\mathrm{with}\quad \tilde{\nu}_f \equiv \nu_f \left[ 1+\left(\frac{q_g}{a_g}\right)^2\right]^{-\frac{1}{2}}.
\end{aligned}
\end{flalign}
It implies that if the $g$-mode is ignored in the analysis but is present in the data, one would infer an effective fundamental mode frequency \textit{smaller} than the true one, which might have important consequences for determining the NS parameters and its dense matter properties. For example, assume $a_g=0.8$ and $q_g=0.1$, which could occur for a $1.4\,M_{\odot}$ hybrid star with a large energy density jump, as shown in Figs. \ref{fig:overlap_integrals} and \ref{jump_frequency}. It follows that $\tilde{\nu}_f=0.99 \nu_f$, which is a $1\%$ difference. 
Assuming heavier NSs ($1.8\,M_{\odot}, 2\,M_{\odot}$), for $n_q/n_h\simeq 1.9$ we have that $a_g=0.65$ and $q_g=0.1$---see also Figs. \ref{fig:overlap_integrals} and \ref{jump_frequency}---, meaning that $\tilde{\nu}_f=0.98\nu_f$ ($2\%$ difference). 
Since the fundamental mode scales with the mean density ($\propto M/R^3$) of the star as $\nu_f\propto (M/R^3)^{\frac{1}{2}}$ \cite{1998MNRAS.299.1059A}, $|\Delta \nu_f/\nu_f|=(3/2)\Delta R/R$ for any given mass $M$. For $a_g=0.8$ and $q_g=0.1$ ($1.4\,M_{\odot}$), $\Delta R/R\sim 0.7\%$; for $a_g=0.65$ and $q_g=0.1$ ($1.8\,M_{\odot}$, $2\,M_{\odot}$), $\Delta R/R\sim 1.3\%$. 

\noindent{\bf {\it Discussion and Conclusions.---}}We found that the discontinuity $g$-mode can compete with the fundamental $f$-mode concerning the overlap integral and other observables (such as the GW phase) for hybrid stars. 
When the stellar mass is near the phase-transition mass,  nonlinearities in $Q_g$ ($\Delta \phi_g$) are observed because it is significantly influenced by the smaller extent of the quark phase and the complex behavior of the $g$-mode eigenfunctions $W$ and $V$ across varying density discontinuities (see the Supplemental Material). Notably, these eigenfunctions can exhibit sign changes within the hadronic phase, thereby markedly affecting $Q_g$, see Eq.~\eqref{overlap_integral}. However, these effects are smoothed out when the stellar mass is further from the phase-transition mass due to the increase of the quark phase and its larger contribution to $Q_g$. 

In our dataset, the point at $n_q/n_h = 1.1$ is the global maximum of $\Delta\phi_g$ within our current resolution. It is followed by smaller local extrema whose amplitudes decrease as $n_q/n_h$ increases; these features arise from nontrivial $g$-mode eigenfunction rearrangements and their progressive smoothing as the quark core grows, rather than from numerical inaccuracies (see the Supplemental Material). Additional small local extrema for $n_q/n_h \lesssim 1.4$ cannot be ruled out with a finer grid and may depend on the equation of state. As $n_q/n_h\to 1$, the density jump—and thus the interfacial restoring force—vanishes, so $\Delta\phi_g$ is expected to approach zero. This trend is consistent with our $1.4M_\odot$ sequence, where $\Delta\phi_g$ at $n_q/n_h=1.075$ is smaller than at $1.1$; it is even clearer in the $1.8M_\odot$ case, for which a less dense $n_q/n_h$ grid shows the same decrease toward the $n_q/n_h\to 1$ limit. A precise exploration of the $n_q/n_h\to1$ regime would require a slow-motion reformulation of the perturbation equations for the near-zero-frequency limit \citep{finn1986} (or more advanced methods such as those presented in \citep{2002PhRvD..66j4002A,2015PhRvD..92f3009K,2025PhRvD.111f3029K}), denser sampling in $n_q/n_h$ and tighter numerical controls (e.g., resolution, tolerances, explicit convergence tests), because the discontinuity $g$-mode frequency tends to zero and the problem becomes numerically delicate. A dedicated study of this asymptotic regime will be pursued in future work.

Nonlinearities in $\Delta\phi_g$ for stars close to the transition mass with small density jumps ($n_q/n_h\approx 1.1$) are observationally relevant, as they imply low resonant frequencies and potentially large GW phase shifts (Fig.~\ref{fig:Number_cycles_variation}). Because the transition mass in our case is close to $1.4\,M_\odot$, one may wonder about GW170817. This event does not rule out phase transitions in NSs because the nondetection of dynamical tides is compatible with (i) a quark--hadron density jump $n_q/n_h \gtrsim 1.1$; (ii) a phase transition mass above the GW170817 component masses; (iii) a supranuclear EOS that differs from the one used here.
Given this astrophysical relevance, further dedicated studies are needed. Another important question is whether linear perturbation theory remains valid at the $g$-mode resonance. Linear theory requires that the ratio between the maximum perturbation amplitude and the radius is much smaller than unity, ${\cal A}_g^{\rm max}\!\ll\! R$. Using Eqs.~(2.6) and (6.3) of \citet{1994MNRAS.270..611L}, we estimate that at the resonance radius  ${\cal A}_g^{\rm max}/R\simeq|a_{g,\rm{max}}||\xi^r_g/R| \lesssim |a_{g,\rm{max}}|\approx 0.24$ for $M=1.4M_{\odot}$, $q=1$, $|Q_g|\sim10^{-1}$ and $\nu_g=500\,\mathrm{Hz}$—the parameters that maximize $\Delta\phi_g$ in Fig.~\ref{fig:Number_cycles_variation}. Since ${\cal A}_g^{\rm max}\propto \nu_g^{-\frac{5}{6}}$ \citep{1994MNRAS.270..611L}, larger frequencies lead to smaller maximum amplitudes. Thus, the discontinuity $g$-mode remains consistent with the linear regime; nevertheless, precise predictions will require incorporating nonlinear effects and relativistic corrections to the Keplerian orbital parameters \citep{2014LRR....17....2B} and the tidal field \citep{2018PhRvD..97l4048P}.

Neglecting the possible contribution of the discontinuity $g$-mode to dynamical tides---particularly when its frequency lies close to that of the $f$-mode, as can occur for large density jumps---may introduce systematic uncertainties in the NS radius. We estimate these uncertainties to be usually smaller than $1\%-2\%$. This might be relevant for 3G GW detectors, as well as for upcoming electromagnetic missions such as eXTP, ATHENA, and STROBE-X. Therefore, stellar structure beyond a single-phase description should not be overlooked in the context of these observations.

Another important aspect that requires further investigation is the efficiency of the mechanisms responsible for dissipating the kinetic energy stored in the $g$-mode. The long damping times associated with it suggest that GW emission alone is insufficient to significantly deplete the $g$-mode energy before merger. However, identifying the dominant dissipation channels—such as neutrino emission and heat generation—remains challenging, as they depend sensitively on the interplay between temperature and density \citep{tonetto2022}. This difficulty is even greater in the quark phase of hybrid stars due to the poorly understood microphysics of dense QCD.

The effect of the EOS on the detectability of the $g$-modes excited during the inspiral phase of compact binaries and the relation of the $g$-mode parameters (frequencies and damping times) with the EOS parameters is still not fully understood; a recent summary, with a special focus on planned 3G GW detectors, is provided by the ET Blue Book, Chap. 6 \citep{2025arXiv250312263A}. The damping time of the $g$-mode is significantly longer than that of the $f$-mode, implying that it emits GWs for longer but less efficiently. We show that despite its weaker emission, the $g$-mode can still induce relevant orbital changes. Given the $g$-mode induced phase shift imprint on the GW signal along its evolution in frequency, one can compare it with the measurement uncertainty, specifically phase calibration errors \cite{PhysRevD.103.042004,Read_2023}. In case of LVK O4 observing run, the phase uncertainty is $<10$ deg ($<0.175$ rad) in the range of 20-2000 Hz \cite{PhysRevD.111.062002}, which makes the detection of a $g$-mode plausible even with current detectors' infrastructure for a sufficiently loud event exhibiting binary components with central densities near the phase transition density (masses close to the mass at which the phase transition occurs), as shown in Fig.~\ref{fig:Number_cycles_variation}. For 3G detectors (ET, CE), the uncertainty will be an order-of-magnitude smaller, ${\approx}1$ deg  
\cite{2020PhRvR...2b3151P}, see also Chap. 10 in \cite{2025arXiv250312263A}.

\noindent{\bf {\it Acknowledgements.---}}J.P.P. acknowledges financial support from CNPq---Conselho Nacional de Desenvolvimento Cient\'ifico e Tecnol\'ogico---under grant No. 174612/2023-0. L.T. acknowledges financial support from INFN, the Italian National Institute for Nuclear Research, 
under grant TEONGRAV. M.B. acknowledges partial support from the Polish National Science Centre grant No. 2021/43/B/ST9/01714. J.L.Z. gratefully acknowledges support from the Polish National Science Centre through grant No. 2018/29/B/ST9/02013.

\bibliography{bibliography}

@ARTICLE{DouchinH2001,
       author = {{Douchin}, F. and {Haensel}, P.},
        title = "{A unified equation of state of dense matter and neutron star structure}",
      journal = {\aap},
         year = "2001",
        month = "Dec",
       volume = {380},
        pages = {151-167},
          doi = {10.1051/0004-6361:20011402},
archivePrefix = {arXiv},
       eprint = {astro-ph/0111092},
 primaryClass = {astro-ph},
       adsurl = {https://ui.adsabs.harvard.edu/abs/2001A&A...380..151D}
}

@ARTICLE{2024PhRvD.109f4004P,
       author = {{Pitre}, Tristan and {Poisson}, Eric},
        title = "{General relativistic dynamical tides in binary inspirals without modes}",
      journal = {\prd},
         year = 2024,
        month = mar,
       volume = {109},
       number = {6},
          eid = {064004},
        pages = {064004},
          doi = {10.1103/PhysRevD.109.064004},
archivePrefix = {arXiv},
       eprint = {2311.04075},
 primaryClass = {gr-qc},
       adsurl = {https://ui.adsabs.harvard.edu/abs/2024PhRvD.109f4004P}
}

@ARTICLE{2021MNRAS.506.2985K,
       author = {{Kuan}, Hao-Jui and {Suvorov}, Arthur G. and {Kokkotas}, Kostas D.},
        title = "{General-relativistic treatment of tidal g-mode resonances in coalescing binaries of neutron stars - I. Theoretical framework and crust breaking}",
      journal = {\mnras},
         year = 2021,
        month = sep,
       volume = {506},
       number = {2},
        pages = {2985-2998},
          doi = {10.1093/mnras/stab1898},
archivePrefix = {arXiv},
       eprint = {2106.16123},
 primaryClass = {gr-qc},
       adsurl = {https://ui.adsabs.harvard.edu/abs/2021MNRAS.506.2985K}
}

@ARTICLE{2021ApJ...918L..28M,
       author = {{Miller}, M.~C. and {Lamb}, F.~K. and {Dittmann}, A.~J. and {Bogdanov}, S. and {Arzoumanian}, Z. and {Gendreau}, K.~C. and {Guillot}, S. and {Ho}, W.~C.~G. and {Lattimer}, J.~M. and {Loewenstein}, M. and {Morsink}, S.~M. and {Ray}, P.~S. and {Wolff}, M.~T. and {Baker}, C.~L. and {Cazeau}, T. and {Manthripragada}, S. and {Markwardt}, C.~B. and {Okajima}, T. and {Pollard}, S. and {Cognard}, I. and {Cromartie}, H.~T. and {Fonseca}, E. and {Guillemot}, L. and {Kerr}, M. and {Parthasarathy}, A. and {Pennucci}, T.~T. and {Ransom}, S. and {Stairs}, I.},
        title = "{The Radius of PSR J0740+6620 from NICER and XMM-Newton Data}",
      journal = {\apjl},
         year = 2021,
        month = sep,
       volume = {918},
       number = {2},
          eid = {L28},
        pages = {L28},
          doi = {10.3847/2041-8213/ac089b},
archivePrefix = {arXiv},
       eprint = {2105.06979},
 primaryClass = {astro-ph.HE},
       adsurl = {https://ui.adsabs.harvard.edu/abs/2021ApJ...918L..28M}
}

@ARTICLE{2020PhRvD.101f3007E,
       author = {{Essick}, Reed and {Landry}, Philippe and {Holz}, Daniel E.},
        title = "{Nonparametric inference of neutron star composition, equation of state, and maximum mass with GW170817}",
      journal = {\prd},
         year = 2020,
        month = mar,
       volume = {101},
       number = {6},
          eid = {063007},
        pages = {063007},
          doi = {10.1103/PhysRevD.101.063007},
archivePrefix = {arXiv},
       eprint = {1910.09740},
 primaryClass = {astro-ph.HE},
       adsurl = {https://ui.adsabs.harvard.edu/abs/2020PhRvD.101f3007E}
}

@ARTICLE{2019PhRvX...9a1001A,
       author = {{Abbott}, B.~P. and {Abbott}, R. and {Abbott}, T.~D. and {Acernese}, F. and
         {Ackley}, K. and {Adams}, C. and {Adams}, T. and {Addesso}, P. and
         {Adhikari}, R.~X. and others},
        title = "{Properties of the Binary Neutron Star Merger GW170817}",
      journal = {Physical Review X},
         year = "2019",
        month = "Jan",
       volume = {9},
       number = {1},
          eid = {011001},
        pages = {011001},
          doi = {10.1103/PhysRevX.9.011001},
archivePrefix = {arXiv},
       eprint = {1805.11579},
 primaryClass = {gr-qc},
       adsurl = {https://ui.adsabs.harvard.edu/abs/2019PhRvX...9a1001A}
}

@ARTICLE{2021MNRAS.503..533A,
       author = {{Andersson}, N. and {Pnigouras}, P.},
        title = "{The phenomenology of dynamical neutron star tides}",
      journal = {\mnras},
         year = 2021,
        month = may,
       volume = {503},
       number = {1},
        pages = {533-539},
          doi = {10.1093/mnras/stab371},
       adsurl = {https://ui.adsabs.harvard.edu/abs/2021MNRAS.503..533A}
}

@ARTICLE{2007A&A...465..533Z,
       author = {{Zdunik}, J.~L. and {Bejger}, M. and {Haensel}, P. and {Gourgoulhon}, E.},
        title = "{Energy release associated with a first-order phase transition in a rotating neutron star core}",
      journal = {\aap},
         year = 2007,
        month = apr,
       volume = {465},
       number = {2},
        pages = {533-539},
          doi = {10.1051/0004-6361:20066515},
archivePrefix = {arXiv},
       eprint = {astro-ph/0610188},
 primaryClass = {astro-ph},
       adsurl = {https://ui.adsabs.harvard.edu/abs/2007A&A...465..533Z}
}

@ARTICLE{2003ApJ...586.1250B,
       author = {{Berezhiani}, Z. and {Bombaci}, I. and {Drago}, A. and {Frontera}, F. and {Lavagno}, A.},
        title = "{Gamma-Ray Bursts from Delayed Collapse of Neutron Stars to Quark Matter Stars}",
      journal = {\apj},
         year = 2003,
        month = apr,
       volume = {586},
       number = {2},
        pages = {1250-1253},
          doi = {10.1086/367756},
archivePrefix = {arXiv},
       eprint = {astro-ph/0209257},
 primaryClass = {astro-ph},
       adsurl = {https://ui.adsabs.harvard.edu/abs/2003ApJ...586.1250B}
}

@ARTICLE{2025JCAP...01..073V,
       author = {{Ventagli}, Giulia and {Saltas}, Ippocratis D.},
        title = "{Deep learning inference of the neutron star equation of state}",
      journal = {\jcap},
         year = 2025,
        month = jan,
       volume = {2025},
       number = {1},
          eid = {073},
        pages = {073},
          doi = {10.1088/1475-7516/2025/01/073},
archivePrefix = {arXiv},
       eprint = {2405.17908},
 primaryClass = {astro-ph.HE},
       adsurl = {https://ui.adsabs.harvard.edu/abs/2025JCAP...01..073V}
}

@ARTICLE{2002PhRvD..66j4002A,
       author = {{Andersson}, N. and {Comer}, G.~L. and {Langlois}, D.},
        title = "{Oscillations of general relativistic superfluid neutron stars}",
      journal = {\prd},
         year = 2002,
        month = nov,
       volume = {66},
       number = {10},
          eid = {104002},
        pages = {104002},
          doi = {10.1103/PhysRevD.66.104002},
archivePrefix = {arXiv},
       eprint = {gr-qc/0203039},
 primaryClass = {gr-qc},
       adsurl = {https://ui.adsabs.harvard.edu/abs/2002PhRvD..66j4002A}
}

@ARTICLE{2022PhRvX..12a1058A,
       author = {{Annala}, Eemeli and {Gorda}, Tyler and {Katerini}, Evangelia and {Kurkela}, Aleksi and {N{\"a}ttil{\"a}}, Joonas and {Paschalidis}, Vasileios and {Vuorinen}, Aleksi},
        title = "{Multimessenger Constraints for Ultradense Matter}",
      journal = {Physical Review X},
         year = 2022,
        month = jan,
       volume = {12},
       number = {1},
          eid = {011058},
        pages = {011058},
          doi = {10.1103/PhysRevX.12.011058},
archivePrefix = {arXiv},
       eprint = {2105.05132},
 primaryClass = {astro-ph.HE},
       adsurl = {https://ui.adsabs.harvard.edu/abs/2022PhRvX..12a1058A}
}

@ARTICLE{2024PhRvD.109f3035C,
       author = {{Christian}, Jan-Erik and {Schaffner-Bielich}, J{\"u}rgen and {Rosswog}, Stephan},
        title = "{Which first order phase transitions to quark matter are possible in neutron stars?}",
      journal = {\prd},
         year = 2024,
        month = mar,
       volume = {109},
       number = {6},
          eid = {063035},
        pages = {063035},
          doi = {10.1103/PhysRevD.109.063035},
archivePrefix = {arXiv},
       eprint = {2312.10148},
 primaryClass = {nucl-th},
       adsurl = {https://ui.adsabs.harvard.edu/abs/2024PhRvD.109f3035C}
}

@ARTICLE{2025PhRvD.111f3007M,
       author = {{Mendes}, Melissa and {Christian}, Jan-Erik and {Fattoyev}, Farrukh J. and {Schaffner-Bielich}, J{\"u}rgen},
        title = "{Constraining twin stars with cold neutron star cooling data}",
      journal = {\prd},
         year = 2025,
        month = mar,
       volume = {111},
       number = {6},
          eid = {063007},
        pages = {063007},
          doi = {10.1103/PhysRevD.111.063007},
archivePrefix = {arXiv},
       eprint = {2408.05287},
 primaryClass = {astro-ph.HE},
       adsurl = {https://ui.adsabs.harvard.edu/abs/2025PhRvD.111f3007M}
}

@ARTICLE{2025PhRvD.111f3029K,
       author = {{Kr{\"u}ger}, Christian J. and {Sotani}, Hajime},
        title = "{Impact of the relativistic Cowling approximation on shear and interface modes of neutron stars}",
      journal = {\prd},
         year = 2025,
        month = mar,
       volume = {111},
       number = {6},
          eid = {063029},
        pages = {063029},
          doi = {10.1103/PhysRevD.111.063029},
archivePrefix = {arXiv},
       eprint = {2411.03940},
 primaryClass = {gr-qc},
       adsurl = {https://ui.adsabs.harvard.edu/abs/2025PhRvD.111f3029K}
}

@ARTICLE{2025ApJ...983...17H,
       author = {{Huang}, Chun and {Sourav}, Shashwat},
        title = "{Constraining First-order Phase Transition inside Neutron Stars with Application of Bayesian Techniques on PSR J0437{\textendash}4715 NICER Data}",
      journal = {\apj},
         year = 2025,
        month = apr,
       volume = {983},
       number = {1},
          eid = {17},
        pages = {17},
          doi = {10.3847/1538-4357/adbb67},
archivePrefix = {arXiv},
       eprint = {2502.11976},
 primaryClass = {astro-ph.HE},
       adsurl = {https://ui.adsabs.harvard.edu/abs/2025ApJ...983...17H}
}

@ARTICLE{2020A&A...642A..78M,
       author = {{Morawski}, F. and {Bejger}, M.},
        title = "{Neural network reconstruction of the dense matter equation of state derived from the parameters of neutron stars}",
      journal = {\aap},
         year = 2020,
        month = oct,
       volume = {642},
          eid = {A78},
        pages = {A78},
          doi = {10.1051/0004-6361/202038130},
archivePrefix = {arXiv},
       eprint = {2006.07194},
 primaryClass = {astro-ph.HE},
       adsurl = {https://ui.adsabs.harvard.edu/abs/2020A&A...642A..78M}
}

@ARTICLE{2023PhRvD.107j3042R,
       author = {{Rau}, Peter B. and {Sedrakian}, Armen},
        title = "{Two first-order phase transitions in hybrid compact stars: Higher-order multiplet stars, reaction modes, and intermediate conversion speeds}",
      journal = {\prd},
         year = 2023,
        month = may,
       volume = {107},
       number = {10},
          eid = {103042},
        pages = {103042},
          doi = {10.1103/PhysRevD.107.103042},
archivePrefix = {arXiv},
       eprint = {2212.09828},
 primaryClass = {astro-ph.HE},
       adsurl = {https://ui.adsabs.harvard.edu/abs/2023PhRvD.107j3042R}
}

@ARTICLE{2020GReGr..52..109C,
       author = {{Chatziioannou}, Katerina},
        title = "{Neutron-star tidal deformability and equation-of-state constraints}",
      journal = {General Relativity and Gravitation},
         year = 2020,
        month = nov,
       volume = {52},
       number = {11},
          eid = {109},
        pages = {109},
          doi = {10.1007/s10714-020-02754-3},
archivePrefix = {arXiv},
       eprint = {2006.03168},
 primaryClass = {gr-qc},
       adsurl = {https://ui.adsabs.harvard.edu/abs/2020GReGr..52..109C}
}

@ARTICLE{2018PhRvL.120z1103M,
       author = {{Most}, Elias R. and {Weih}, Lukas R. and {Rezzolla}, Luciano and {Schaffner-Bielich}, J{\"u}rgen},
        title = "{New Constraints on Radii and Tidal Deformabilities of Neutron Stars from GW170817}",
      journal = {\prl},
         year = 2018,
        month = jun,
       volume = {120},
       number = {26},
          eid = {261103},
        pages = {261103},
          doi = {10.1103/PhysRevLett.120.261103},
archivePrefix = {arXiv},
       eprint = {1803.00549},
 primaryClass = {gr-qc},
       adsurl = {https://ui.adsabs.harvard.edu/abs/2018PhRvL.120z1103M}
}

@ARTICLE{2020PhRvL.124q1103W,
       author = {{Weih}, Lukas R. and {Hanauske}, Matthias and {Rezzolla}, Luciano},
        title = "{Postmerger Gravitational-Wave Signatures of Phase Transitions in Binary Mergers}",
      journal = {\prl},
         year = 2020,
        month = may,
       volume = {124},
       number = {17},
          eid = {171103},
        pages = {171103},
          doi = {10.1103/PhysRevLett.124.171103},
archivePrefix = {arXiv},
       eprint = {1912.09340},
 primaryClass = {gr-qc},
       adsurl = {https://ui.adsabs.harvard.edu/abs/2020PhRvL.124q1103W}
}

@ARTICLE{2019PhRvL.122f1102B,
       author = {{Bauswein}, Andreas and {Bastian}, Niels-Uwe F. and {Blaschke}, David B. and {Chatziioannou}, Katerina and {Clark}, James A. and {Fischer}, Tobias and {Oertel}, Micaela},
        title = "{Identifying a First-Order Phase Transition in Neutron-Star Mergers through Gravitational Waves}",
      journal = {\prl},
         year = 2019,
        month = feb,
       volume = {122},
       number = {6},
          eid = {061102},
        pages = {061102},
          doi = {10.1103/PhysRevLett.122.061102},
archivePrefix = {arXiv},
       eprint = {1809.01116},
 primaryClass = {astro-ph.HE},
       adsurl = {https://ui.adsabs.harvard.edu/abs/2019PhRvL.122f1102B}
}

@ARTICLE{2022PhRvD.105l3032W,
       author = {{Williams}, Natalie and {Pratten}, Geraint and {Schmidt}, Patricia},
        title = "{Prospects for distinguishing dynamical tides in inspiralling binary neutron stars with third generation gravitational-wave detectors}",
      journal = {\prd},
         year = 2022,
        month = jun,
       volume = {105},
       number = {12},
          eid = {123032},
        pages = {123032},
          doi = {10.1103/PhysRevD.105.123032},
archivePrefix = {arXiv},
       eprint = {2203.00623},
 primaryClass = {astro-ph.HE},
       adsurl = {https://ui.adsabs.harvard.edu/abs/2022PhRvD.105l3032W}
}

@ARTICLE{2024ApJ...971L..19R,
       author = {{Rutherford}, Nathan and {Mendes}, Melissa and {Svensson}, Isak and {Schwenk}, Achim and {Watts}, Anna L. and {Hebeler}, Kai and {Keller}, Jonas and {Prescod-Weinstein}, Chanda and {Choudhury}, Devarshi and {Raaijmakers}, Geert and {Salmi}, Tuomo and {Timmerman}, Patrick and {Vinciguerra}, Serena and {Guillot}, Sebastien and {Lattimer}, James M.},
        title = "{Constraining the Dense Matter Equation of State with New NICER Mass{\textendash}Radius Measurements and New Chiral Effective Field Theory Inputs}",
      journal = {\apjl},
         year = 2024,
        month = aug,
       volume = {971},
       number = {1},
          eid = {L19},
        pages = {L19},
          doi = {10.3847/2041-8213/ad5f02},
archivePrefix = {arXiv},
       eprint = {2407.06790},
 primaryClass = {astro-ph.HE},
       adsurl = {https://ui.adsabs.harvard.edu/abs/2024ApJ...971L..19R}
}

@ARTICLE{2022PhRvL.129h1102P,
       author = {{Pratten}, Geraint and {Schmidt}, Patricia and {Williams}, Natalie},
        title = "{Impact of Dynamical Tides on the Reconstruction of the Neutron Star Equation of State}",
      journal = {\prl},
         year = 2022,
        month = aug,
       volume = {129},
       number = {8},
          eid = {081102},
        pages = {081102},
          doi = {10.1103/PhysRevLett.129.081102},
archivePrefix = {arXiv},
       eprint = {2109.07566},
 primaryClass = {astro-ph.HE},
       adsurl = {https://ui.adsabs.harvard.edu/abs/2022PhRvL.129h1102P}
}

@ARTICLE{2023PhRvD.108b3010R,
       author = {{Raithel}, Carolyn A. and {Most}, Elias R.},
        title = "{Tidal deformability doppelg{\"a}nger: Implications of a low-density phase transition in the neutron star equation of state}",
      journal = {\prd},
         year = 2023,
        month = jul,
       volume = {108},
       number = {2},
          eid = {023010},
        pages = {023010},
          doi = {10.1103/PhysRevD.108.023010},
archivePrefix = {arXiv},
       eprint = {2208.04295},
 primaryClass = {astro-ph.HE},
       adsurl = {https://ui.adsabs.harvard.edu/abs/2023PhRvD.108b3010R}
}

@ARTICLE{2003MNRAS.338..389M,
       author = {{Miniutti}, G. and {Pons}, J.~A. and {Berti}, E. and {Gualtieri}, L. and {Ferrari}, V.},
        title = "{Non-radial oscillation modes as a probe of density discontinuities in neutron stars}",
      journal = {\mnras},
         year = 2003,
        month = jan,
       volume = {338},
       number = {2},
        pages = {389-400},
          doi = {10.1046/j.1365-8711.2003.06057.x},
archivePrefix = {arXiv},
       eprint = {astro-ph/0206142},
 primaryClass = {astro-ph},
       adsurl = {https://ui.adsabs.harvard.edu/abs/2003MNRAS.338..389M}
}

@ARTICLE{2023PhRvL.130t1403R,
       author = {{Raithel}, Carolyn A. and {Most}, Elias R.},
        title = "{Degeneracy in the Inference of Phase Transitions in the Neutron Star Equation of State from Gravitational Wave Data}",
      journal = {\prl},
         year = 2023,
        month = may,
       volume = {130},
       number = {20},
          eid = {201403},
        pages = {201403},
          doi = {10.1103/PhysRevLett.130.201403},
archivePrefix = {arXiv},
       eprint = {2208.04294},
 primaryClass = {astro-ph.HE},
       adsurl = {https://ui.adsabs.harvard.edu/abs/2023PhRvL.130t1403R}
}

@ARTICLE{2019PhRvD.100b1501S,
       author = {{Schmidt}, Patricia and {Hinderer}, Tanja},
        title = "{Frequency domain model of f -mode dynamic tides in gravitational waveforms from compact binary inspirals}",
      journal = {\prd},
         year = 2019,
        month = jul,
       volume = {100},
       number = {2},
          eid = {021501},
        pages = {021501},
          doi = {10.1103/PhysRevD.100.021501},
archivePrefix = {arXiv},
       eprint = {1905.00818},
 primaryClass = {gr-qc},
       adsurl = {https://ui.adsabs.harvard.edu/abs/2019PhRvD.100b1501S}
}

@ARTICLE{2008A&A...479..515Z,
   author = {{Zdunik}, J.~L. and {Bejger}, M. and {Haensel}, P. and {Gourgoulhon}, E.
	},
    title = "{Strong first-order phase transition in a rotating neutron star core and the associated energy release}",
  journal = {\aap},
archivePrefix = "arXiv",
   eprint = {0707.3691},
     year = 2008,
    month = feb,
   volume = 479,
    pages = {515-522},
      doi = {10.1051/0004-6361:20078346},
   adsurl = {http://adsabs.harvard.edu/abs/2008A%26A...479..515Z}
}

@ARTICLE{2020PASA...37...47A,
       author = {{Ackley}, K. and {Adya}, V.~B. and {Agrawal}, P. and {Altin}, P. and {Ashton}, G. and {Bailes}, M. and {Baltinas}, E. and {Barbuio} and et al.},
        title = "{Neutron Star Extreme Matter Observatory: A kilohertz-band gravitational-wave detector in the global network}",
      journal = {\pasa},
         year = 2020,
        month = nov,
       volume = {37},
          eid = {e047},
        pages = {e047},
          doi = {10.1017/pasa.2020.39},
archivePrefix = {arXiv},
       eprint = {2007.03128},
 primaryClass = {astro-ph.HE},
       adsurl = {https://ui.adsabs.harvard.edu/abs/2020PASA...37...47A}
}

@ARTICLE{2021arXiv210506980R,
        author = {{Riley}, Thomas E. and {Watts}, Anna L. and {Ray}, Paul S. and {Bogdanov}, Slavko and {Guillot}, Sebastien and {Morsink}, Sharon M. and {Bilous}, Anna V. and {Arzoumanian}, Zaven and {Choudhury}, Devarshi and {Deneva}, Julia S. and {Gendreau}, Keith C. and {Harding}, Alice K. and {Ho}, Wynn C.~G. and {Lattimer}, James M. and {Loewenstein}, Michael and {Ludlam}, Renee M. and {Markwardt}, Craig B. and {Okajima}, Takashi and {Prescod-Weinstein}, Chanda and {Remillard}, Ronald A. and {Wolff}, Michael T. and {Fonseca}, Emmanuel and {Cromartie}, H. Thankful and {Kerr}, Matthew and {Pennucci}, Timothy T. and {Parthasarathy}, Aditya and {Ransom}, Scott and {Stairs}, Ingrid and {Guillemot}, Lucas and {Cognard}, Ismael},
        title = "{A NICER View of the Massive Pulsar PSR J0740+6620 Informed by Radio Timing and XMM-Newton Spectroscopy}",
      journal = {\apjl},
         year = 2021,
        month = sep,
       volume = {918},
       number = {2},
          eid = {L27},
        pages = {L27},
          doi = {10.3847/2041-8213/ac0a81},
archivePrefix = {arXiv},
       eprint = {2105.06980},
 primaryClass = {astro-ph.HE},
       adsurl = {https://ui.adsabs.harvard.edu/abs/2021ApJ...918L..27R}
}

@ARTICLE{2021arXiv210506981R,
       author = {{Raaijmakers}, G. and {Greif}, S.~K. and {Hebeler}, K. and {Hinderer}, T. and {Nissanke}, S. and {Schwenk}, A. and {Riley}, T.~E. and {Watts}, A.~L. and {Lattimer}, J.~M. and {Ho}, W.~C.~G.},
        title = "{Constraints on the dense matter equation of state and neutron star properties from NICER's mass-radius estimate of PSR J0740+6620 and multimessenger observations}",
      journal = {arXiv e-prints},
         year = 2021,
        month = may,
          eid = {arXiv:2105.06981},
        pages = {arXiv:2105.06981},
archivePrefix = {arXiv},
       eprint = {2105.06981},
 primaryClass = {astro-ph.HE},
       adsurl = {https://ui.adsabs.harvard.edu/abs/2021arXiv210506981R}
}

@ARTICLE{2019ApJ...887L..24M,
       author = {{Miller}, M.~C. and {Lamb}, F.~K. and {Dittmann}, A.~J. and {Bogdanov}, S. and {Arzoumanian}, Z. and {Gendreau}, K.~C. and {Guillot}, S. and {Harding}, A.~K. and {Ho}, W.~C.~G. and {Lattimer}, J.~M. and {Ludlam}, R.~M. and {Mahmoodifar}, S. and {Morsink}, S.~M. and {Ray}, P.~S. and {Strohmayer}, T.~E. and {Wood}, K.~S. and {Enoto}, T. and {Foster}, R. and {Okajima}, T. and {Prigozhin}, G. and {Soong}, Y.},
        title = "{PSR J0030+0451 Mass and Radius from NICER Data and Implications for the Properties of Neutron Star Matter}",
      journal = {\apjl},
         year = 2019,
        month = dec,
       volume = {887},
       number = {1},
          eid = {L24},
        pages = {L24},
          doi = {10.3847/2041-8213/ab50c5},
archivePrefix = {arXiv},
       eprint = {1912.05705},
 primaryClass = {astro-ph.HE},
       adsurl = {https://ui.adsabs.harvard.edu/abs/2019ApJ...887L..24M}
}

@ARTICLE{2019ApJ...887L..21R,
       author = {{Riley}, T.~E. and {Watts}, A.~L. and {Bogdanov}, S. and {Ray}, P.~S. and {Ludlam}, R.~M. and {Guillot}, S. and {Arzoumanian}, Z. and {Baker}, C.~L. and {Bilous}, A.~V. and {Chakrabarty}, D. and {Gendreau}, K.~C. and {Harding}, A.~K. and {Ho}, W.~C.~G. and {Lattimer}, J.~M. and {Morsink}, S.~M. and {Strohmayer}, T.~E.},
        title = "{A NICER View of PSR J0030+0451: Millisecond Pulsar Parameter Estimation}",
      journal = {\apjl},
         year = 2019,
        month = dec,
       volume = {887},
       number = {1},
          eid = {L21},
        pages = {L21},
          doi = {10.3847/2041-8213/ab481c},
archivePrefix = {arXiv},
       eprint = {1912.05702},
 primaryClass = {astro-ph.HE},
       adsurl = {https://ui.adsabs.harvard.edu/abs/2019ApJ...887L..21R}
}

@ARTICLE{2021PhRvD.103f3015L,
       author = {{Lau}, Shu Yan and {Yagi}, Kent},
        title = "{Probing hybrid stars with gravitational waves via interfacial modes}",
      journal = {\prd},
         year = 2021,
        month = mar,
       volume = {103},
       number = {6},
          eid = {063015},
        pages = {063015},
          doi = {10.1103/PhysRevD.103.063015},
archivePrefix = {arXiv},
       eprint = {2012.13000},
 primaryClass = {astro-ph.HE},
       adsurl = {https://ui.adsabs.harvard.edu/abs/2021PhRvD.103f3015L}
}

@ARTICLE{2018ApJ...860...12P,
       author = {{Pereira}, Jonas P. and {Flores}, C{\'e}sar V. and {Lugones}, Germ{\'a}n},
        title = "{Phase Transition Effects on the Dynamical Stability of Hybrid Neutron Stars}",
      journal = {\apj},
         year = 2018,
        month = jun,
       volume = {860},
       number = {1},
          eid = {12},
        pages = {12},
          doi = {10.3847/1538-4357/aabfbf},
archivePrefix = {arXiv},
       eprint = {1706.09371},
 primaryClass = {gr-qc},
       adsurl = {https://ui.adsabs.harvard.edu/abs/2018ApJ...860...12P}
}

@ARTICLE{2012PhRvL.108a1102T,
       author = {{Tsang}, David and {Read}, Jocelyn S. and {Hinderer}, Tanja and {Piro}, Anthony L. and {Bondarescu}, Ruxandra},
        title = "{Resonant Shattering of Neutron Star Crusts}",
      journal = {\prl},
         year = 2012,
        month = jan,
       volume = {108},
       number = {1},
          eid = {011102},
        pages = {011102},
          doi = {10.1103/PhysRevLett.108.011102},
archivePrefix = {arXiv},
       eprint = {1110.0467},
 primaryClass = {astro-ph.HE},
       adsurl = {https://ui.adsabs.harvard.edu/abs/2012PhRvL.108a1102T}
}

@ARTICLE{2014LRR....17....2B,
       author = {{Blanchet}, Luc},
        title = "{Gravitational Radiation from Post-Newtonian Sources and Inspiralling Compact Binaries}",
      journal = {Living Reviews in Relativity},
         year = 2014,
        month = dec,
       volume = {17},
       number = {1},
          eid = {2},
        pages = {2},
          doi = {10.12942/lrr-2014-2},
archivePrefix = {arXiv},
       eprint = {1310.1528},
 primaryClass = {gr-qc},
       adsurl = {https://ui.adsabs.harvard.edu/abs/2014LRR....17....2B}
}

@ARTICLE{2018PhRvD..97l4048P,
       author = {{Poisson}, Eric and {Corrigan}, Eamonn},
        title = "{Nonrotating black hole in a post-Newtonian tidal environment. II.}",
      journal = {\prd},
         year = 2018,
        month = jun,
       volume = {97},
       number = {12},
          eid = {124048},
        pages = {124048},
          doi = {10.1103/PhysRevD.97.124048},
archivePrefix = {arXiv},
       eprint = {1804.01848},
 primaryClass = {gr-qc},
       adsurl = {https://ui.adsabs.harvard.edu/abs/2018PhRvD..97l4048P}
}

@ARTICLE{2020NatPh..16..907A,
       author = {{Annala}, Eemeli and {Gorda}, Tyler and {Kurkela}, Aleksi and {N{\"a}ttil{\"a}}, Joonas and {Vuorinen}, Aleksi},
        title = "{Evidence for quark-matter cores in massive neutron stars}",
      journal = {Nature Physics},
         year = 2020,
        month = jun,
       volume = {16},
       number = {9},
        pages = {907-910},
          doi = {10.1038/s41567-020-0914-9},
archivePrefix = {arXiv},
       eprint = {1903.09121},
 primaryClass = {astro-ph.HE},
       adsurl = {https://ui.adsabs.harvard.edu/abs/2020NatPh..16..907A}
}

@ARTICLE{2018PhRvL.121p1101A,
       author = {{Abbott}, B.~P. and {Abbott}, R. and {Abbott}, T.~D. and {Acernese}, F. and {Ackley}, K. and {Adams}, C. and {Adams}, T. and {Addesso}, P. and {Adhikari}, R.~X. and {Adya}, V.~B. and et al.},
        title = "{GW170817: Measurements of Neutron Star Radii and Equation of State}",
      journal = {\prl},
         year = 2018,
        month = oct,
       volume = {121},
       number = {16},
          eid = {161101},
        pages = {161101},
          doi = {10.1103/PhysRevLett.121.161101},
archivePrefix = {arXiv},
       eprint = {1805.11581},
 primaryClass = {gr-qc},
       adsurl = {https://ui.adsabs.harvard.edu/abs/2018PhRvL.121p1101A}
}

@ARTICLE{1990MNRAS.245...82F,
       author = {{Finn}, Lee Samuel},
        title = "{Non-radial pulsations of neutron stars with a crust}",
      journal = {\mnras},
         year = 1990,
        month = jul,
       volume = {245},
        pages = {82},
          doi = {10.1093/mnras/245.1.82},
       adsurl = {https://ui.adsabs.harvard.edu/abs/1990MNRAS.245...82F}
}

@ARTICLE{1987MNRAS.227..265F,
       author = {{Finn}, Lee Samuel},
        title = "{G-modes in zero-temperature neutron stars}",
      journal = {\mnras},
         year = 1987,
        month = jul,
       volume = {227},
        pages = {265-293},
          doi = {10.1093/mnras/227.2.265},
       adsurl = {https://ui.adsabs.harvard.edu/abs/1987MNRAS.227..265F}
}

@ARTICLE{2020PhRvD.101h3002S,
       author = {{Suvorov}, Arthur G. and {Kokkotas}, Kostas D.},
        title = "{Precursor flares of short gamma-ray bursts from crust yielding due to tidal resonances in coalescing binaries of rotating, magnetized neutron stars}",
      journal = {\prd},
         year = 2020,
        month = apr,
       volume = {101},
       number = {8},
          eid = {083002},
        pages = {083002},
          doi = {10.1103/PhysRevD.101.083002},
archivePrefix = {arXiv},
       eprint = {2003.05673},
 primaryClass = {astro-ph.HE},
       adsurl = {https://ui.adsabs.harvard.edu/abs/2020PhRvD.101h3002S}
}

@ARTICLE{2021ApJ...910..145P,
       author = {{Pereira}, Jonas P. and {Bejger}, Micha{\l} and {Tonetto}, Lucas and {Lugones}, Germ{\'a}n and {Haensel}, Pawe{\l} and {Zdunik}, Julian Leszek and {Sieniawska}, Magdalena},
        title = "{Probing Elastic Quark Phases in Hybrid Stars with Radius Measurements}",
      journal = {\apj},
         year = 2021,
        month = apr,
       volume = {910},
       number = {2},
          eid = {145},
        pages = {145},
          doi = {10.3847/1538-4357/abe633},
archivePrefix = {arXiv},
       eprint = {2011.06361},
 primaryClass = {astro-ph.HE},
       adsurl = {https://ui.adsabs.harvard.edu/abs/2021ApJ...910..145P}
}

@ARTICLE{2021MNRAS.508.1732K,
       author = {{Kuan}, Hao-Jui and {Suvorov}, Arthur G. and {Kokkotas}, Kostas D.},
        title = "{General-relativistic treatment of tidal g-mode resonances in coalescing binaries of neutron stars - II. As triggers for precursor flares of short gamma-ray bursts}",
      journal = {\mnras},
         year = 2021,
        month = dec,
       volume = {508},
       number = {2},
        pages = {1732-1744},
          doi = {10.1093/mnras/stab2658},
archivePrefix = {arXiv},
       eprint = {2107.00533},
 primaryClass = {astro-ph.HE},
       adsurl = {https://ui.adsabs.harvard.edu/abs/2021MNRAS.508.1732K}
}

@ARTICLE{2022MNRAS.513.4045K,
       author = {{Kuan}, Hao-Jui and {Kr{\"u}ger}, Christian J. and {Suvorov}, Arthur G. and {Kokkotas}, Kostas D.},
        title = "{Constraining equation-of-state groups from g-mode asteroseismology}",
      journal = {\mnras},
         year = 2022,
        month = jul,
       volume = {513},
       number = {3},
        pages = {4045-4056},
          doi = {10.1093/mnras/stac1101},
archivePrefix = {arXiv},
       eprint = {2204.08492},
 primaryClass = {gr-qc},
       adsurl = {https://ui.adsabs.harvard.edu/abs/2022MNRAS.513.4045K}
}

@ARTICLE{2022MNRAS.514.1628K,
       author = {{Kerin}, A.~D. and {Melatos}, A.},
        title = "{Mountain formation by repeated, inhomogeneous crustal failure in a neutron star}",
      journal = {\mnras},
         year = 2022,
        month = aug,
       volume = {514},
       number = {2},
        pages = {1628-1644},
          doi = {10.1093/mnras/stac1351},
archivePrefix = {arXiv},
       eprint = {2205.15026},
 primaryClass = {astro-ph.HE},
       adsurl = {https://ui.adsabs.harvard.edu/abs/2022MNRAS.514.1628K}
}

@ARTICLE{2019A&A...622A.174S,
       author = {{Sieniawska}, M. and {Turcza{\'n}ski}, W. and {Bejger}, M. and
         {Zdunik}, J.~L.},
        title = "{Tidal deformability and other global parameters of compact stars with strong phase transitions}",
      journal = {\aap},
         year = "2019",
        month = "Feb",
       volume = {622},
          eid = {A174},
        pages = {A174},
          doi = {10.1051/0004-6361/201833969},
archivePrefix = {arXiv},
       eprint = {1807.11581},
 primaryClass = {astro-ph.HE},
       adsurl = {https://ui.adsabs.harvard.edu/abs/2019A&A...622A.174S}
}

@ARTICLE{2004gr.qc....11025B,
       author = {{Berti}, Emanuele},
        title = "{Black hole quasinormal modes: hints of quantum gravity?}",
      journal = {arXiv e-prints},
         year = 2004,
        month = nov,
          eid = {gr-qc/0411025},
        pages = {gr-qc/0411025},
          doi = {10.48550/arXiv.gr-qc/0411025},
archivePrefix = {arXiv},
       eprint = {gr-qc/0411025},
 primaryClass = {gr-qc},
       adsurl = {https://ui.adsabs.harvard.edu/abs/2004gr.qc....11025B}
}

@ARTICLE{2001PhRvD..65b4010S,
       author = {{Sotani}, Hajime and {Tominaga}, Kazuhiro and {Maeda}, Kei-Ichi},
        title = "{Density discontinuity of a neutron star and gravitational waves}",
      journal = {\prd},
         year = 2001,
        month = dec,
       volume = {65},
       number = {2},
          eid = {024010},
        pages = {024010},
          doi = {10.1103/PhysRevD.65.024010},
archivePrefix = {arXiv},
       eprint = {gr-qc/0108060},
 primaryClass = {gr-qc},
       adsurl = {https://ui.adsabs.harvard.edu/abs/2001PhRvD..65b4010S}
}

@ARTICLE{2020PhRvD.101l3029T,
       author = {{Tonetto}, L. and {Lugones}, G.},
        title = "{Discontinuity gravity modes in hybrid stars: Assessing the role of rapid and slow phase conversions}",
      journal = {\prd},
         year = 2020,
        month = jun,
       volume = {101},
       number = {12},
          eid = {123029},
        pages = {123029},
          doi = {10.1103/PhysRevD.101.123029},
archivePrefix = {arXiv},
       eprint = {2003.01259},
 primaryClass = {astro-ph.HE},
       adsurl = {https://ui.adsabs.harvard.edu/abs/2020PhRvD.101l3029T}
}

@ARTICLE{2021MNRAS.504.1273P,
       author = {{Passamonti}, A. and {Andersson}, N. and {Pnigouras}, P.},
        title = "{Dynamical tides in neutron stars: the impact of the crust}",
      journal = {\mnras},
         year = 2021,
        month = jun,
       volume = {504},
       number = {1},
        pages = {1273-1293},
          doi = {10.1093/mnras/stab870},
archivePrefix = {arXiv},
       eprint = {2012.09637},
 primaryClass = {astro-ph.HE},
       adsurl = {https://ui.adsabs.harvard.edu/abs/2021MNRAS.504.1273P}
}

@ARTICLE{1998MNRAS.299.1059A,
       author = {{Andersson}, Nils and {Kokkotas}, Kostas D.},
        title = "{Towards gravitational wave asteroseismology}",
      journal = {\mnras},
         year = 1998,
        month = oct,
       volume = {299},
       number = {4},
        pages = {1059-1068},
          doi = {10.1046/j.1365-8711.1998.01840.x},
archivePrefix = {arXiv},
       eprint = {gr-qc/9711088},
 primaryClass = {gr-qc},
       adsurl = {https://ui.adsabs.harvard.edu/abs/1998MNRAS.299.1059A}
}

@ARTICLE{1994MNRAS.270..611L,
       author = {{Lai}, D.},
        title = "{Resonant Oscillations and Tidal Heating in Coalescing Binary Neutron Stars}",
      journal = {\mnras},
         year = 1994,
        month = oct,
       volume = {270},
        pages = {611},
          doi = {10.1093/mnras/270.3.611},
archivePrefix = {arXiv},
       eprint = {astro-ph/9404062},
 primaryClass = {astro-ph},
       adsurl = {https://ui.adsabs.harvard.edu/abs/1994MNRAS.270..611L}
}

@ARTICLE{2019SCPMA..6229503W,
       author = {{Watts}, Anna L. and {Yu}, WenFei and {Poutanen}, Juri and {Zhang}, Shu and {Bhattacharyya} and et al.},
        title = "{Dense matter with eXTP}",
      journal = {Science China Physics, Mechanics, and Astronomy},
         year = 2019,
        month = feb,
       volume = {62},
       number = {2},
          eid = {29503},
        pages = {29503},
          doi = {10.1007/s11433-017-9188-4},
archivePrefix = {arXiv},
       eprint = {1812.04021},
 primaryClass = {astro-ph.HE},
       adsurl = {https://ui.adsabs.harvard.edu/abs/2019SCPMA..6229503W}
}

@ARTICLE{2021ApJ...913...27L,
       author = {{Li}, Ang and {Miao}, Zhiqiang and {Han}, Sophia and {Zhang}, Bing},
        title = "{Constraints on the Maximum Mass of Neutron Stars with a Quark Core from GW170817 and NICER PSR J0030+0451 Data}",
      journal = {\apj},
         year = 2021,
        month = may,
       volume = {913},
       number = {1},
          eid = {27},
        pages = {27},
          doi = {10.3847/1538-4357/abf355},
archivePrefix = {arXiv},
       eprint = {2103.15119},
 primaryClass = {astro-ph.HE},
       adsurl = {https://ui.adsabs.harvard.edu/abs/2021ApJ...913...27L}
}

@ARTICLE{2021PhRvL.126q2502A,
       author = {{PREX Collaboration}},
        title = "{Accurate Determination of the Neutron Skin Thickness of $^{208}$Pb through Parity-Violation in Electron Scattering}",
      journal = {\prl},
         year = 2021,
        month = apr,
       volume = {126},
       number = {17},
          eid = {172502},
        pages = {172502},
          doi = {10.1103/PhysRevLett.126.172502},
archivePrefix = {arXiv},
       eprint = {2102.10767},
 primaryClass = {nucl-ex},
       adsurl = {https://ui.adsabs.harvard.edu/abs/2021PhRvL.126q2502A}
}

@ARTICLE{2021arXiv210813071J,
       author = {{Jie Li}, Jia and {Sedrakian}, Armen and {Alford}, Mark},
        title = "{Relativistic hybrid stars in light of the NICER PSR J0740+6620 radius measurement}",
      journal = {arXiv e-prints},
         year = 2021,
        month = aug,
          eid = {arXiv:2108.13071},
        pages = {arXiv:2108.13071},
archivePrefix = {arXiv},
       eprint = {2108.13071},
 primaryClass = {astro-ph.HE},
       adsurl = {https://ui.adsabs.harvard.edu/abs/2021arXiv210813071J}
}

@ARTICLE{2021PhRvL.126q2503R,
       author = {{Reed}, Brendan T. and {Fattoyev}, F.~J. and {Horowitz}, C.~J. and {Piekarewicz}, J.},
        title = "{Implications of PREX-2 on the Equation of State of Neutron-Rich Matter}",
      journal = {\prl},
         year = 2021,
        month = apr,
       volume = {126},
       number = {17},
          eid = {172503},
        pages = {172503},
          doi = {10.1103/PhysRevLett.126.172503},
archivePrefix = {arXiv},
       eprint = {2101.03193},
 primaryClass = {nucl-th},
       adsurl = {https://ui.adsabs.harvard.edu/abs/2021PhRvL.126q2503R}
}

@ARTICLE{2013arXiv1306.2307N,
       author = {{Nandra}, K. and others},
        title = "{The Hot and Energetic Universe: A White Paper presenting the science theme motivating the Athena+ mission}",
      journal = {arXiv e-prints},
         year = 2013,
        month = jun,
          eid = {arXiv:1306.2307},
        pages = {arXiv:1306.2307},
archivePrefix = {arXiv},
       eprint = {1306.2307},
 primaryClass = {astro-ph.HE},
       adsurl = {https://ui.adsabs.harvard.edu/abs/2013arXiv1306.2307N}
}

@inproceedings{NICER,
author = {Keith C. Gendreau and Zaven Arzoumanian and Takashi Okajima},
title = {{The Neutron star Interior Composition ExploreR (NICER): an Explorer mission of opportunity for soft x-ray timing spectroscopy}},
volume = {8443},
booktitle = {Space Telescopes and Instrumentation 2012: Ultraviolet to Gamma Ray},
editor = {Tadayuki Takahashi and Stephen S. Murray and Jan-Willem A. den Herder},
organization = {International Society for Optics and Photonics},
publisher = {SPIE},
pages = {322 -- 329},
year = {2012},
doi = {},
URL = {https://doi.org/10.1117/12.926396}
}

@ARTICLE{2015CQGra..32g4001L,
       author = {{Aasi}, J. and others},
        title = "{Advanced LIGO}",
      journal = {Classical and Quantum Gravity},
         year = 2015,
        month = apr,
       volume = {32},
       number = {7},
          eid = {074001},
        pages = {074001},
          doi = {10.1088/0264-9381/32/7/074001},
archivePrefix = {arXiv},
       eprint = {1411.4547},
 primaryClass = {gr-qc},
       adsurl = {https://ui.adsabs.harvard.edu/abs/2015CQGra..32g4001L}
}

@ARTICLE{2015CQGra..32b4001A,
       author = {{Acernese}, F. and others}, 
        title = "{Advanced Virgo: a second-generation interferometric gravitational wave detector}",
      journal = {Classical and Quantum Gravity},
         year = 2015,
        month = jan,
       volume = {32},
       number = {2},
          eid = {024001},
        pages = {024001},
          doi = {10.1088/0264-9381/32/2/024001},
archivePrefix = {arXiv},
       eprint = {1408.3978},
 primaryClass = {gr-qc},
       adsurl = {https://ui.adsabs.harvard.edu/abs/2015CQGra..32b4001A}
}

@ARTICLE{2019NatAs...3...35K,
       author = {{Akutsu}, T. and others},
        title = "{KAGRA: 2.5 generation interferometric gravitational wave detector}",
      journal = {Nature Astronomy},
         year = 2019,
        month = jan,
       volume = {3},
        pages = {35-40},
          doi = {10.1038/s41550-018-0658-y},
archivePrefix = {arXiv},
       eprint = {1811.08079},
 primaryClass = {gr-qc},
       adsurl = {https://ui.adsabs.harvard.edu/abs/2019NatAs...3...35K}
}

@ARTICLE{2020JCAP...03..050M,
       author = {{Maggiore}, Michele and {Van Den Broeck}, Chris and {Bartolo}, Nicola and {Belgacem}, Enis and {Bertacca}, Daniele and {Bizouard}, Marie Anne and {Branchesi}, Marica and {Clesse}, Sebastien and {Foffa}, Stefano and {Garc{\'\i}a-Bellido}, Juan and {Grimm}, Stefan and {Harms}, Jan and {Hinderer}, Tanja and {Matarrese}, Sabino and {Palomba}, Cristiano and {Peloso}, Marco and {Ricciardone}, Angelo and {Sakellariadou}, Mairi},
        title = "{Science case for the Einstein telescope}",
      journal = {\jcap},
         year = 2020,
        month = mar,
       volume = {2020},
       number = {3},
          eid = {050},
        pages = {050},
          doi = {10.1088/1475-7516/2020/03/050},
archivePrefix = {arXiv},
       eprint = {1912.02622},
 primaryClass = {astro-ph.CO},
       adsurl = {https://ui.adsabs.harvard.edu/abs/2020JCAP...03..050M}
}

@ARTICLE{2015PhRvD..92f3009K,
       author = {{Kr{\"u}ger}, C.~J. and {Ho}, W.~C.~G. and {Andersson}, N.},
        title = "{Seismology of adolescent neutron stars: Accounting for thermal effects and crust elasticity}",
      journal = {\prd},
         year = 2015,
        month = sep,
       volume = {92},
       number = {6},
          eid = {063009},
        pages = {063009},
          doi = {10.1103/PhysRevD.92.063009},
archivePrefix = {arXiv},
       eprint = {1402.5656},
 primaryClass = {gr-qc},
       adsurl = {https://ui.adsabs.harvard.edu/abs/2015PhRvD..92f3009K}
}

@INPROCEEDINGS{2019BAAS...51g..35R,
       author = {{Reitze}, David and {Adhikari}, Rana X. and {Ballmer}, Stefan and {Barish}, Barry and {Barsotti}, Lisa and {Billingsley}, GariLynn and {Brown}, Duncan A. and {Chen}, Yanbei and {Coyne}, Dennis and {Eisenstein}, Robert and {Evans}, Matthew and {Fritschel}, Peter and {Hall}, Evan D. and {Lazzarini}, Albert and {Lovelace}, Geoffrey and {Read}, Jocelyn and {Sathyaprakash}, B.~S. and {Shoemaker}, David and {Smith}, Joshua and {Torrie}, Calum and {Vitale}, Salvatore and {Weiss}, Rainer and {Wipf}, Christopher and {Zucker}, Michael},
        title = "{Cosmic Explorer: The U.S. Contribution to Gravitational-Wave Astronomy beyond LIGO}",
    booktitle = {Bulletin of the American Astronomical Society},
         year = 2019,
       volume = {51},
        month = sep,
          eid = {35},
        pages = {35},
archivePrefix = {arXiv},
       eprint = {1907.04833},
 primaryClass = {astro-ph.IM},
       adsurl = {https://ui.adsabs.harvard.edu/abs/2019BAAS...51g..35R}
}

@ARTICLE{2017PhRvL.119p1101A,
       author = {{LIGO Scientific Collaboration} and {Virgo Collaboration}},
        title = "{GW170817: Observation of Gravitational Waves from a Binary Neutron Star Inspiral}",
      journal = {\prl},
         year = 2017,
        month = oct,
       volume = {119},
       number = {16},
          eid = {161101},
        pages = {161101},
          doi = {10.1103/PhysRevLett.119.161101},
archivePrefix = {arXiv},
       eprint = {1710.05832},
 primaryClass = {gr-qc},
       adsurl = {https://ui.adsabs.harvard.edu/abs/2017PhRvL.119p1101A}
}

@ARTICLE{2019arXiv190303035R,
       author = {{Ray}, Paul S. and {Arzoumanian}, Zaven and {Ballantyne}, David and {Bozzo}, Enrico and {Brandt}, Soren and {Brenneman}, Laura and {Chakrabarty}, Deepto and {Christophersen}, Marc and others},
        title = "{STROBE-X: X-ray Timing and Spectroscopy on Dynamical Timescales from Microseconds to Years}",
      journal = {arXiv e-prints},
         year = 2019,
        month = mar,
          eid = {arXiv:1903.03035},
        pages = {arXiv:1903.03035},
archivePrefix = {arXiv},
       eprint = {1903.03035},
 primaryClass = {astro-ph.IM},
       adsurl = {https://ui.adsabs.harvard.edu/abs/2019arXiv190303035R}
}

@ARTICLE{2008RvMP...80.1455A,
       author = {{Alford}, Mark G. and {Schmitt}, Andreas and {Rajagopal}, Krishna and {Sch{\"a}fer}, Thomas},
        title = "{Color superconductivity in dense quark matter}",
      journal = {Reviews of Modern Physics},
         year = 2008,
        month = oct,
       volume = {80},
       number = {4},
        pages = {1455-1515},
          doi = {10.1103/RevModPhys.80.1455},
archivePrefix = {arXiv},
       eprint = {0709.4635},
 primaryClass = {hep-ph},
       adsurl = {https://ui.adsabs.harvard.edu/abs/2008RvMP...80.1455A}
}

@ARTICLE{2021PhRvC.103c5802X,
       author = {{Xie}, Wen-Jie and {Li}, Bao-An},
        title = "{Bayesian inference of the dense-matter equation of state encapsulating a first-order hadron-quark phase transition from observables of canonical neutron stars}",
      journal = {\prc},
         year = 2021,
        month = mar,
       volume = {103},
       number = {3},
          eid = {035802},
        pages = {035802},
          doi = {10.1103/PhysRevC.103.035802},
archivePrefix = {arXiv},
       eprint = {2009.13653},
 primaryClass = {nucl-th},
       adsurl = {https://ui.adsabs.harvard.edu/abs/2021PhRvC.103c5802X}
}

@ARTICLE{2021arXiv210812368C,
       author = {{Chatziioannou}, Katerina},
        title = "{Uncertainty limits on neutron star radius measurements with gravitational waves}",
      journal = {arXiv e-prints},
         year = 2021,
        month = aug,
          eid = {arXiv:2108.12368},
        pages = {arXiv:2108.12368},
archivePrefix = {arXiv},
       eprint = {2108.12368},
 primaryClass = {gr-qc},
       adsurl = {https://ui.adsabs.harvard.edu/abs/2021arXiv210812368C}
}

@article{PhysRevD.98.124014,
  title = {Impact of high-order tidal terms on binary neutron-star waveforms},
  author = {Jim\'enez Forteza, Xisco and Abdelsalhin, Tiziano and Pani, Paolo and Gualtieri, Leonardo},
  journal = {Phys. Rev. D},
  volume = {98},
  issue = {12},
  pages = {124014},
  numpages = {16},
  year = {2018},
  month = {Dec},
  publisher = {American Physical Society},
  doi = {10.1103/PhysRevD.98.124014},
  url = {https://link.aps.org/doi/10.1103/PhysRevD.98.124014}
}

@INPROCEEDINGS{2016SPIE.9905E..1QZ,
   author = {{Zhang}, S.~N. and {Feroci}, M. and {Santangelo}, A. and {Dong}, Y.~W. and 
	{Feng}, H. and {Lu}, F.~J. and {Nandra}, K. and et al.
	},
    title = "{eXTP: Enhanced X-ray Timing and Polarization mission}",
booktitle = {Space Telescopes and Instrumentation 2016: Ultraviolet to Gamma Ray},
     year = 2016,
   series = {Proceedings of the SPIE},
   volume = 9905,
archivePrefix = "arXiv",
   eprint = {1607.08823},
 primaryClass = "astro-ph.IM",
    month = jul,
      eid = {99051Q},
    pages = {99051Q},
      doi = {10.1117/12.2232034},
   adsurl = {http://adsabs.harvard.edu/abs/2016SPIE.9905E..1QZ}
}

@ARTICLE{2024PhRvD.110d3013W,
       author = {{Walker}, Kris and {Smith}, Rory and {Thrane}, Eric and {Reardon}, Daniel J.},
        title = "{Precision constraints on the neutron star equation of state with third-generation gravitational-wave observatories}",
      journal = {\prd},
         year = 2024,
        month = aug,
       volume = {110},
       number = {4},
          eid = {043013},
        pages = {043013},
          doi = {10.1103/PhysRevD.110.043013},
archivePrefix = {arXiv},
       eprint = {2401.02604},
 primaryClass = {astro-ph.HE},
       adsurl = {https://ui.adsabs.harvard.edu/abs/2024PhRvD.110d3013W}
}

@ARTICLE{2025arXiv250406181C,
       author = {{Counsell}, A.~R. and {Gittins}, F. and {Andersson}, N. and {Tews}, I.},
        title = "{Interface modes in inspiralling neutron stars: A gravitational-wave probe of first-order phase transitions}",
      journal = {arXiv e-prints},
         year = 2025,
        month = apr,
          eid = {arXiv:2504.06181},
        pages = {arXiv:2504.06181},
archivePrefix = {arXiv},
       eprint = {2504.06181},
 primaryClass = {gr-qc},
       adsurl = {https://ui.adsabs.harvard.edu/abs/2025arXiv250406181C}
}

@ARTICLE{2024ApJ...964...31M,
       author = {{Miao}, Zhiqiang and {Zhou}, Enping and {Li}, Ang},
        title = "{Resolving Phase Transition Properties of Dense Matter through Tidal-excited g-mode from Inspiralling Neutron Stars}",
      journal = {\apj},
         year = 2024,
        month = mar,
       volume = {964},
       number = {1},
          eid = {31},
        pages = {31},
          doi = {10.3847/1538-4357/ad27cd},
archivePrefix = {arXiv},
       eprint = {2305.08401},
 primaryClass = {nucl-th},
       adsurl = {https://ui.adsabs.harvard.edu/abs/2024ApJ...964...31M}
}

@ARTICLE{2020arXiv200310781P,
       author = {{Pereira}, Jonas P. and {Bejger}, Micha{\l} and {Andersson}, Nils and
         {Gittins}, Fabian},
        title = "{Tidal Deformations of Hybrid Stars with Sharp Phase Transitions and Elastic Crusts}",
      journal = {\apj},
         year = 2020,
        month = may,
       volume = {895},
       number = {1},
          eid = {28},
        pages = {28},
          doi = {10.3847/1538-4357/ab8aca},
archivePrefix = {arXiv},
       eprint = {2003.10781},
 primaryClass = {gr-qc},
       adsurl = {https://ui.adsabs.harvard.edu/abs/2020ApJ...895...28P}
}

@ARTICLE{thornecampolattaro1967,
   author = {Thorne, K.S. and Campolattaro, A.},
    title = {Non-Radial Pulsation of General-Relativistic Stellar Models. I. Analytic Analysis for $l \geq 2$},
  journal = {Astrophys. J.},
     year = {1967},
    month = {September},
   volume = {149},
    pages = {591},
      doi = {10.1086/149288},
}

@ARTICLE{thornecampolattaro1967erratum,
   author = {{Thorne}, K.~S. and {Campolattaro}, A.},
    title = "{Erratum: Non-Radial Pulsation of General-Relativistivc Stellar Models. I. Analytic Analysis for $l \geq$ 2}",
  journal = {Astrophys. J.},
     year = 1968,
    month = may,
   volume = 152,
    pages = {673},
      doi = {10.1086/149586},
}

@ARTICLE{lindblomdetweiler1983,
   author = {{Lindblom}, L. and {Detweiler}, S.~L.},
    title = "{The quadrupole oscillations of neutron stars}",
  journal = {Astrophys. J. Supp.},
     year = 1983,
    month = sep,
   volume = 53,
    pages = {73-92},
      doi = {10.1086/190884},
   adsurl = {http://adsabs.harvard.edu/abs/1983ApJS...53...73L}
}

@ARTICLE{lindblomdetweiler1985,
   author = {{Detweiler}, S. and {Lindblom}, L.},
    title = "{On the nonradial pulsations of general relativistic stellar models}",
  journal = {Astrophys. J. },
     year = 1985,
    month = may,
   volume = 292,
    pages = {12-15},
      doi = {10.1086/163127},
   adsurl = {http://adsabs.harvard.edu/abs/1985ApJ...292...12D}
}

@article{reggewheeler1957,
  title = {Stability of a Schwarzschild Singularity},
  author = {Regge, Tullio and Wheeler, John A.},
  journal = {Phys. Rev.},
  volume = {108},
  issue = {4},
  pages = {1063--1069},
  numpages = {0},
  year = {1957},
  month = {Nov},
  publisher = {American Physical Society},
  doi = {10.1103/PhysRev.108.1063},
  url = {https://link.aps.org/doi/10.1103/PhysRev.108.1063}
}

@article{nollert1993,
  title = {Quasinormal modes of Schwarzschild black holes: The determination of quasinormal frequencies with very large imaginary parts},
  author = {Nollert, Hans-Peter},
  journal = {Phys. Rev. D},
  volume = {47},
  issue = {12},
  pages = {5253--5258},
  numpages = {0},
  year = {1993},
  month = {Jun},
  publisher = {American Physical Society},
  doi = {10.1103/PhysRevD.47.5253},
  url = {https://link.aps.org/doi/10.1103/PhysRevD.47.5253}
}

@article{leins1993,
  title = {Nonradial oscillations of neutron stars: A new branch of strongly damped normal modes},
  author = {Leins, M. and Nollert, H. -P. and Soffel, M. H.},
  journal = {Phys. Rev. D},
  volume = {48},
  issue = {8},
  pages = {3467--3472},
  numpages = {0},
  year = {1993},
  month = {Oct},
  publisher = {American Physical Society},
  doi = {10.1103/PhysRevD.48.3467},
  url = {https://link.aps.org/doi/10.1103/PhysRevD.48.3467}
}

@article{sotani2022_CF,
  title={Accuracy of one-dimensional approximation in neutron star quasi-normal modes},
  author={Sotani, Hajime},
  journal={The European Physical Journal C},
  volume={82},
  number={5},
  pages={477},
  year={2022},
  publisher={Springer}
}

@article{mariani2022,
    author = {Mariani, Mauro and Tonetto, Lucas and Rodríguez, M Camila and Celi, Marcos O and Ranea-Sandoval, Ignacio F and Orsaria, Milva G and Pérez Martínez, Aurora},
    title = {Oscillating magnetized hybrid stars under the magnifying glass of multimessenger observations},
    journal = {Monthly Notices of the Royal Astronomical Society},
    volume = {512},
    number = {1},
    pages = {517-534},
    year = {2022},
    month = {02},
    issn = {0035-8711},
    doi = {10.1093/mnras/stac546},
    url = {https://doi.org/10.1093/mnras/stac546},
    eprint = {https://academic.oup.com/mnras/article-pdf/512/1/517/42901285/stac546.pdf},
}

@article{finn1986,
    author = {Finn, Lee Samuel},
    title = {g-modes of non-radially pulsating relativistic stars: the slow-motion formalism},
    journal = {Monthly Notices of the Royal Astronomical Society},
    volume = {222},
    number = {3},
    pages = {393-416},
    year = {1986},
    month = {10},
    issn = {0035-8711},
    doi = {10.1093/mnras/222.3.393},
    url = {https://doi.org/10.1093/mnras/222.3.393},
    eprint = {https://academic.oup.com/mnras/article-pdf/222/3/393/3226144/mnras222-0393.pdf},
}

@ARTICLE{2025arXiv250312263A,
       author = {{Abac}, A. and others},
        title = "{The Science of the Einstein Telescope}",
      journal = {arXiv e-prints},
         year = 2025,
        month = mar,
          eid = {arXiv:2503.12263},
        pages = {arXiv:2503.12263},
          doi = {10.48550/arXiv.2503.12263},
archivePrefix = {arXiv},
       eprint = {2503.12263},
 primaryClass = {gr-qc},
       adsurl = {https://ui.adsabs.harvard.edu/abs/2025arXiv250312263A}
}

@ARTICLE{2020CQGra..37p5003A,
       author = {{Adhikari}, R.~X. and others},
        title = "{A cryogenic silicon interferometer for gravitational-wave detection}",
      journal = {Classical and Quantum Gravity},
         year = 2020,
        month = aug,
       volume = {37},
       number = {16},
          eid = {165003},
        pages = {165003},
          doi = {10.1088/1361-6382/ab9143},
archivePrefix = {arXiv},
       eprint = {2001.11173},
 primaryClass = {astro-ph.IM},
       adsurl = {https://ui.adsabs.harvard.edu/abs/2020CQGra..37p5003A}
}

@article{Read_2023,
doi = {10.1088/1361-6382/acd29b},
url = {https://dx.doi.org/10.1088/1361-6382/acd29b},
year = {2023},
month = {may},
publisher = {IOP Publishing},
volume = {40},
number = {13},
pages = {135002},
author = {Read, Jocelyn},
title = {Waveform uncertainty quantification and interpretation for gravitational-wave astronomy},
journal = {Classical and Quantum Gravity},
}

@article{PhysRevD.103.042004,
  title = {Constraining unmodeled physics with compact binary mergers from GWTC-1},
  author = {Edelman, Bruce and Rivera-Paleo, F. J. and Merritt, J. D. and Farr, Ben and Doctor, Zoheyr and Brink, Jeandrew and Farr, Will M. and Gair, Jonathan and Key, Joey Shapiro and McIver, Jess and Nielsen, Alex B.},
  journal = {Phys. Rev. D},
  volume = {103},
  issue = {4},
  pages = {042004},
  numpages = {13},
  year = {2021},
  month = {Feb},
  publisher = {American Physical Society},
  doi = {10.1103/PhysRevD.103.042004},
  url = {https://link.aps.org/doi/10.1103/PhysRevD.103.042004}
}

@article{PhysRevD.111.062002,
  title = {Advanced LIGO detector performance in the fourth observing run},
  author = {Capote, E. and others},
  journal = {Phys. Rev. D},
  volume = {111},
  issue = {6},
  pages = {062002},
  numpages = {30},
  year = {2025},
  month = {Mar},
  publisher = {American Physical Society},
  doi = {10.1103/PhysRevD.111.062002},
  url = {https://link.aps.org/doi/10.1103/PhysRevD.111.062002}
}

@ARTICLE{2020PhRvR...2b3151P,
       author = {{P{\"u}rrer}, Michael and {Haster}, Carl-Johan},
        title = "{Gravitational waveform accuracy requirements for future ground-based detectors}",
      journal = {Physical Review Research},
         year = 2020,
        month = may,
       volume = {2},
       number = {2},
          eid = {023151},
        pages = {023151},
          doi = {10.1103/PhysRevResearch.2.023151},
archivePrefix = {arXiv},
       eprint = {1912.10055},
 primaryClass = {gr-qc},
       adsurl = {https://ui.adsabs.harvard.edu/abs/2020PhRvR...2b3151P}
}

@article{tonetto2022,
  title = {Thermal effects on nuclear matter properties},
  author = {Tonetto, Lucas and Benhar, Omar},
  journal = {Phys. Rev. D},
  volume = {106},
  issue = {10},
  pages = {103020},
  numpages = {11},
  year = {2022},
  month = {Nov},
  publisher = {American Physical Society},
  doi = {10.1103/PhysRevD.106.103020},
  url = {https://link.aps.org/doi/10.1103/PhysRevD.106.103020}
}

\section{Supplemental Material}
\section{Set of equations of state}
For a hybrid star exhibiting a sharp phase transition, two separate equations of state (EOSs) are required: one describing the hadronic phase and another for the quark phase. Below, we outline the methodology used for their construction. For an in-depth discussion, refer to \cite{2021ApJ...910..145P}.
 
A unified EOS for NSs was developed using the SLy4 effective interaction model \citep{DouchinH2001}. This approach provides a consistent physical description of the structure and EOS for both the crust and the core, including the transition region between them, all derived from the same nuclear effective interaction \citep{DouchinH2001}.

To describe the construction of EOSs for the hadronic sector, the SLy4 EOS is connected to a relativistic polytrope represented by $p=\kappa_{\rm ef} n_b^{\gamma}$, with the energy density given by $\rho = p/(\gamma - 1) + n_b m_b$, where $n_b$ denotes the baryon density. This connection begins at a baryon density $n_0$ and extends to a higher density, $n_1$. 
For this study, the parameters $n_0=0.21$ fm$^{-3}$, $\gamma=4.5$ and $n_1=0.335$ fm$^{-3}$ (slightly above $2\rho_{\rm sat}$) are adopted. These parameters are motivated by nuclear theory, which is reliable up to densities around nuclear saturation, and by astrophysical considerations, which suggest that the onset of quark matter may occur above approximately 2 times the nuclear saturation density (see, e.g., \citep{2020NatPh..16..907A,2022PhRvX..12a1058A,2024PhRvD.109f3035C,2025PhRvD.111f3007M}).
The constants $m_b$ ({\it baryon mass} in the polytrope phase) and $\kappa_{\rm ef}$ are determined to ensure continuity of both the pressure and the chemical potential at $n_0$.

The quark phase is modeled using the simplified MIT bag EOS, $p = c_s^2(\rho - \rho_{\star})$, which is matched to the previously described hadronic EOS.  Here, $c_s^2=1$ is chosen to investigate the characteristics of stiff quark matter, in agreement with the Bayesian inference of NSs with phase transitions (hybrid stars) that satisfy observational constraints \citep{2021PhRvC.103c5802X}. (In the future, with multiple NS observations, it will also be possible to better constrain $c_s^2$ and the EOS of neutron stars using neural networks \citep{2020A&A...642A..78M,2025JCAP...01..073V}.) The baryon density jump at the interface between the hadronic and quark phases is treated as a free parameter, with chemical and mechanical equilibrium conditions determining both the magnitude of the density jump and $\rho_{\star}$. For additional details, refer to \cite{2019A&A...622A.174S,2020arXiv200310781P}. 

For (more realistic) quark-phase models with a lower speed of sound ($c_s^2<1$) and fixed phase transition pressure $p_t$, the discontinuity-driven $g$-mode GW phase shift $\Delta\phi_g$ may increase. This is because at a fixed $p_t$, a softer EOS makes the quark core more compressible, enlarging the density jump $\Delta\epsilon$ at the interface and thereby raising the Brunt--Väisälä frequency and the $g$-mode frequency $\nu_g$. Although a larger $\nu_g$ shifts the resonance to later in the inspiral, a stronger discontinuity may enhance the tidal coupling—via a larger buoyancy force—and thereby increase $\Delta\phi_g$. However, because the $g$-mode eigenfrequencies also depend nontrivially on the quark-core size and on $\Delta\epsilon$, the net effect on $Q_g$—and thus on $\Delta\phi_g$—is not straightforward in general and should be established by explicit calculations. We leave them for future work.


In addition to the $M$–$R$ relation shown in Fig.~1 of the main text, we further illustrate the structure of hybrid stars described by the EOSs above by presenting, in Fig.~\ref{fig:R_q_R_ratio}, the ratio of the quark core radius ($R_q$) to the total stellar radius ($R$). This is shown for hybrid stars with fixed masses ($1.4\,M_{\odot}$, $1.8\,M_{\odot}$, and $2.0\,M_{\odot}$) and varying quark-to-hadron baryon number density ratios ($n_q/n_h$). (The relationship between the relative energy density jump $\Delta \epsilon / \epsilon_h\equiv \epsilon_q/\epsilon_h-1$ and the relative baryon number density jump $\Delta n / n_h$ is given by $\Delta \epsilon / \epsilon_h = \left(1 + p_{\rm{t}} / \epsilon_h\right) \Delta n / n_h$, where $p_{\rm{t}}$ is the phase transition pressure.) Note that more massive stars exhibit larger values of $R_q/R$, as expected from their higher central densities.

Assuming the hadronic EOS is reasonably well constrained, there is a practical upper bound on the density jump \(n_q/n_h\): larger jumps yield more compact stars and lower maximum masses, risking tension with observational limits on \(M_{\rm max}\) and radii. Independent inferences place the onset of deconfined quark matter at \(\gtrsim 2\,n_{\rm sat}\) \citep{2020NatPh..16..907A,2022PhRvX..12a1058A,2025PhRvD.111f3007M}, favoring \(n_q/n_h \lesssim 2\) \citep{2024PhRvD.109f3035C,2025PhRvD.111f3007M}. Accordingly, we restrict to \(n_q/n_h \le 1.9\), since larger jumps drive the maximum neutron-star mass below observational bounds for our chosen hadronic EOS. A stiffer hadronic EOS—still consistent with multimessenger constraints—could allow somewhat larger \(n_q/n_h\), but our aim is to span weak and strong transitions. In this sense, the range \(n_q/n_h \le 1.9\) already captures the relevant phenomenology and the conceptual scope of this work. Our working range \(n_q/n_h \le 1.9\) is therefore consistent with both the expected onset of quark matter and the stability of observed neutron stars.

\begin{figure}
\includegraphics[width=\columnwidth]{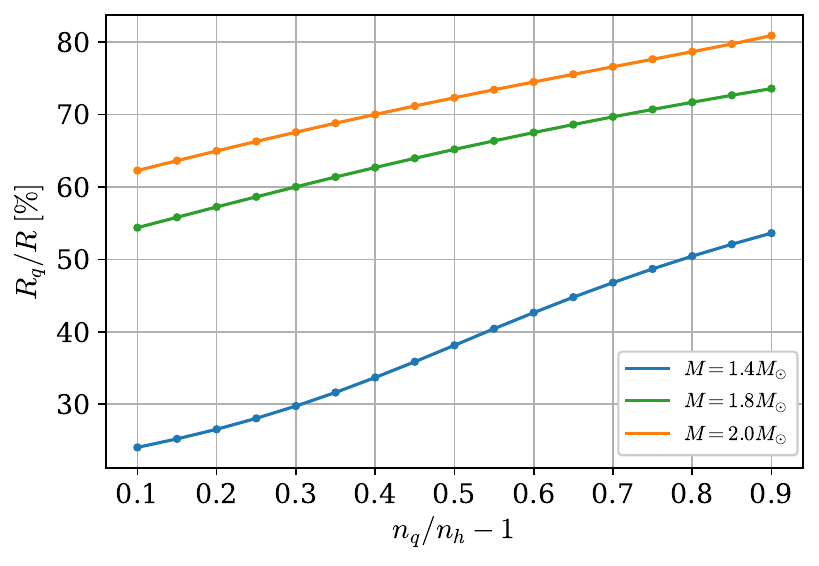}
\caption{Quark core radius $R_q$ relative to the stellar radius $R$ for hybrid stars with fixed masses ($1.4\,M_{\odot}$, $1.8\,M_{\odot}$, and $2.0\,M_{\odot}$), shown as a function of $n_q/n_h$. A gradual increase in $R_q/R$ with $n_q/n_h$ is observed. As expected, more massive stars exhibit relatively larger quark cores compared to their total radii.
\label{fig:R_q_R_ratio}}
\end{figure}

\vskip 1em
\section{Formalism of nonradial modes}
To describe nonradial modes, we follow the standard approach \citep{thornecampolattaro1967,thornecampolattaro1967erratum,lindblomdetweiler1983,lindblomdetweiler1985}. If the background spacetime is stationary and spherically symmetric, 
\begin{equation}
\label{dsz_tov}
ds^{2}=-e^{2\nu(r)} dt^{2} + e^{2\lambda(r)} dr^{2} + r^{2}(d\theta^{2}+\sin^{2}{\theta}d\phi^{2}),
\end{equation} 
the corresponding TOV hydrostatic equilibrium equations are   
\begin{align}
    \frac{dp}{dr} & = -\frac{\epsilon m}{r^2} \left( 1+\frac{p}{\epsilon} \right) \left( 1+ \frac{4 \pi p r^3}{m} \right) \left( 1- \frac{2m}{r} \right)^{-1}, \nonumber \\    
    \frac{dm}{dr} & =4 \pi r^2 \epsilon\quad\mathrm{and}\quad\frac{d \nu}{dr} = - \frac{1}{(\epsilon+p)} \frac{dp}{dr},
\label{eq:TOV}
\end{align}
where $p$ is the pressure, $\epsilon$ is the mass-energy density, and $m(r)$ is the gravitational mass inside sphere of radius $r$ related to the metric function $\lambda(r)$ as 
$e^{\lambda(r)}=1/\sqrt{1-{2m(r)}/{r}}$. The boundary conditions are given by $m(r{=}0)=0$, $p(r{=}R)=0$ and $\nu(r{=}R)=0.5\ln\left(1-2M/R\right)$, where $R$ corresponds to the stellar radius and $M=m(R)$ to the stellar mass.

For the small-amplitude non-radial oscillations, we work in the Regge-Wheeler gauge \citep{reggewheeler1957}, applying it to even-parity\footnote{Odd-parity motions do not emit GWs, therefore we will not take them into consideration.} perturbations, and focusing on the dominating $l=2$ case. 
We connect the calculations of eigenfunctions in the stellar interior to the exterior using the results and notation of \cite{2020PhRvD.101l3029T}. The Lagrangian 3-vector fluid displacement is $\mathbf{\xi} = (\xi_r,\xi_\theta,\xi_\phi)e^{i\omega t}$, with
\begin{eqnarray}
 \xi^r&=&r^{l-1} e^{-\lambda} W(r) Y^l_m, \label{eq:Xi} \\
 \xi^\theta&=&-r^{l-2} V(r) \partial_\theta Y^l_m, \\
 \xi^\phi&=&-r^l (r \ \sin\theta)^{-2} V(r) \partial_\phi Y^l_m,
\label{eq:lagrangian_3vector}
\end{eqnarray}
while the perturbed line element becomes \citep{lindblomdetweiler1985}
\begin{equation}
\begin{aligned}
 ds^{2}= &-e^{2\nu} (1+r^l H_0 Y^l_m e^{i\omega t}) dt^{2}  \\
         & - 2i\omega r^{l+1} H_1 Y^l_m e^{i\omega t} dtdr \\
         & + e^{2\lambda}(1-r^l H_0 Y^l_m e^{i\omega t}) dr^2  \\
         & + r^{2}(1-r^l K Y^l_m e^{i\omega t})(d\theta^{2}+\sin^{2}{\theta}d\phi^{2}).
\label{eq:metric_DL}
\end{aligned}
\end{equation} 
Introducing the variable $X = -r^{-l} e^\nu \Delta p$, with $\Delta p$ being the Lagrangian pressure change, written as:
\begin{equation}
    X = \omega^2 (\epsilon+p) e^{-\nu} V - r^{-1} p,_r e^{(\nu-\lambda)} W + \frac{1}{2} (\epsilon+p) e^{\nu} H_0,
    \label{eq:definitionX}
\end{equation}
where $,_r$ indicates differentiation with respect to $r$, one can write a
fourth-order system of linear equations for $\left(H_{1}, K, W, X\right)$ \citep{lindblomdetweiler1983,lindblomdetweiler1985}:
\begin{eqnarray}
\label{eq:osc_system}
K,_r &=& H_0/r + \tfrac{1}{2}l(l+1)r^{-1}H_1 - [(l+1)/r-\nu,_r]K  \nonumber \\
&& - 8 \pi (\epsilon+p) e^{\lambda} r^{-1} W, \label{eq:oscK} 
\end{eqnarray}
\begin{eqnarray}
H_1,_r &=& -r^{-1} [l+1+2M e^{2\lambda}/r + 4 \pi r^2 e^{2\lambda} (p-\epsilon)] H_1 \nonumber \\
&& + r^{-1} e^{2 \lambda} [H_0 + K - 16 \pi (\epsilon+p) V], \label{eq:oscH1} 
\end{eqnarray}
\begin{eqnarray}
W,_r &=& -(l+1)r^{-1}W + r e^{\lambda} [(\gamma p)^{-1} e^{-\nu} X \nonumber \\
&& - l(l+1)r^{-2} V + \tfrac{1}{2} H_0 + K], \label{eq:oscW}
\end{eqnarray}
\begin{eqnarray}
X,_r &= & -l r^{-1} X + (\epsilon+p)e^{\nu} \{ \tfrac{1}{2} (r^{-1}-\nu,_r) H_0 \nonumber \\
&& + \tfrac{1}{2} [r \omega^2 e^{-2\nu}  + \tfrac{1}{2} l(l+1)/r] H_1 + \tfrac{1}{2} (3 \nu,_r - r^{-1})K  \nonumber \\
&&- l(l+1) \nu,_r r^{-2} V - r^{-1} [4 \pi (\epsilon+p) e^{\lambda}   \nonumber \\
&& + \omega^2 e^{\lambda-2\nu}- r^2 (r^{-2}e^{-\lambda} \nu,_r),_r] W \} \label{eq:oscX}.
\end{eqnarray}
where $\gamma$ is the adabatic index.
As usual, the numerical values of the complex quasi-normal frequency $\omega = \omega_R + i \omega_I$ are obtained imposing a purely outgoing wave behavior at infinity to the Zerilli function \citep{2020PhRvD.101l3029T}. It is assumed that $\omega_I \ll \omega_R$, so that it is reasonable to assume the input trial frequency as a purely real value.

The presence of a sharp phase transition adds boundary conditions to the perturbation problem (for early references on this issue, see \citep{1987MNRAS.227..265F,1990MNRAS.245...82F}).
We assume the case when the phase conversions are slow \citep{2018ApJ...860...12P,2020PhRvD.101l3029T} (implying, among other things, $[W]^+_-=0$), which are appropriate for cold stars \citep{2021ApJ...910..145P}. For any quantity \(A\equiv A(r,t)\), \([A]_{-}^{+}\) denotes its jump across the phase-transition radius \(R_{\rm pt}\), i.e., \(
[A]_{-}^{+} \coloneqq \lim_{q \to 0^{+}} \big[A\big]_{R_{\rm pt}-q}^{R_{\rm pt}+q}.
\)

It follows that almost all perturbation variables are continuous, the only exceptions being $\delta p$ and $V$. Since $[\epsilon]^+_-\neq 0$ and $[p_{,r}]^+_-\neq 0$ (on account of the TOV equations, see the end section for the formalism of nonradial modes) and $[\Delta p]^+_-=0$ for a first-order phase transition, it follows in general that $[\delta p]^+_-=[\Delta p -p_{,r}\xi^r]^+_-= -\xi^r[p_{,r}]^+_-\neq 0$. From the definition of $X$, $[\Delta p]^+_-=0$ implies $[X]^+_-=0$, which, due to Eq.\eqref{eq:definitionX}, leads to $[V]^+_-\neq 0$ in general. The above boundary conditions are essential to ensure the self-consistency of the perturbative treatment.

\vskip 1em
\section{Solution strategy for nonradial modes}
We have solved the system of equations for nonradial perturbations using two methods: (i) Lindblom's standard method \citep{lindblomdetweiler1983,lindblomdetweiler1985} and the method of continued fractions (CF) \citep{2004gr.qc....11025B,nollert1993,leins1993}. 

Lindblom's method is known to perform well for modes where the imaginary part of the frequency satisfies $\omega_i \ll \omega_r$, with $\omega = \omega_r + i \omega_i$. However, due to the boundary conditions imposed at infinity, convergence becomes problematic when $\omega_i$ is too small. For standard stratification $g$-modes, this issue often renders Lindblom's method less effective \citep{2020PhRvD.101l3029T,mariani2022}, and alternative approaches are typically preferred.

In contrast, for the quark-hadron discontinuity $g$-mode examined in this work, the imaginary part $\omega_i$ is larger than that found in standard stratification $g$-modes. This mitigates the convergence issues and simplifies the numerical computation. Nevertheless, the calculation of $\omega_i$ remains the most uncertain part of the process and should be interpreted only as an order-of-magnitude estimate. Future work should refine this analysis, for instance by employing more advanced methods such as those presented in \citep{2002PhRvD..66j4002A,2025PhRvD.111f3029K}.

While Lindblom's method provides sufficiently accurate results for the real part of the frequency, $\omega_r$, we adopt the CF method for a more precise determination. As an initial step, Lindblom’s algorithm is used to estimate the approximate location of $\omega_r$ in frequency space. We then sample the CF expression \citep{2004gr.qc....11025B,nollert1993,leins1993} around this estimate and identify the frequency corresponding to the minimum of the function, as prescribed in \cite{sotani2022_CF}. By definition, the eigenfrequency is the root of the CF expansion coming from a recurrence relation. As shown in \cite{sotani2022_CF}, a sufficiently refined frequency grid allows the minimum point to be taken as a good approximation to the eigenfrequency.

The CF method is well suited for handling both real and complex frequencies and remains robust regardless of the magnitude of $\omega_i$. For the $f$- and $g$-modes considered in this work, we confirm that $\omega_i \ll \omega_r$. Additionally, by comparing the imaginary and real components of the eigenfunctions, we find that the contribution from the imaginary part is significantly smaller than that of the real part. This observation justifies restricting our sampling to the real frequency axis and performing the calculation of the overlap integrals under this assumption. In test cases where the full complex eigenfrequency was computed, we further verified that the imaginary part remains much smaller than the real part. Therefore, for computational efficiency, we limit our analysis to real frequencies.

Once the eigenfrequency is determined using the continued fractions method, we compute the corresponding eigenfunctions and proceed with the evaluation of the overlap integrals and subsequent analyses.

\begin{figure}[!htb]
\includegraphics[width=\columnwidth]{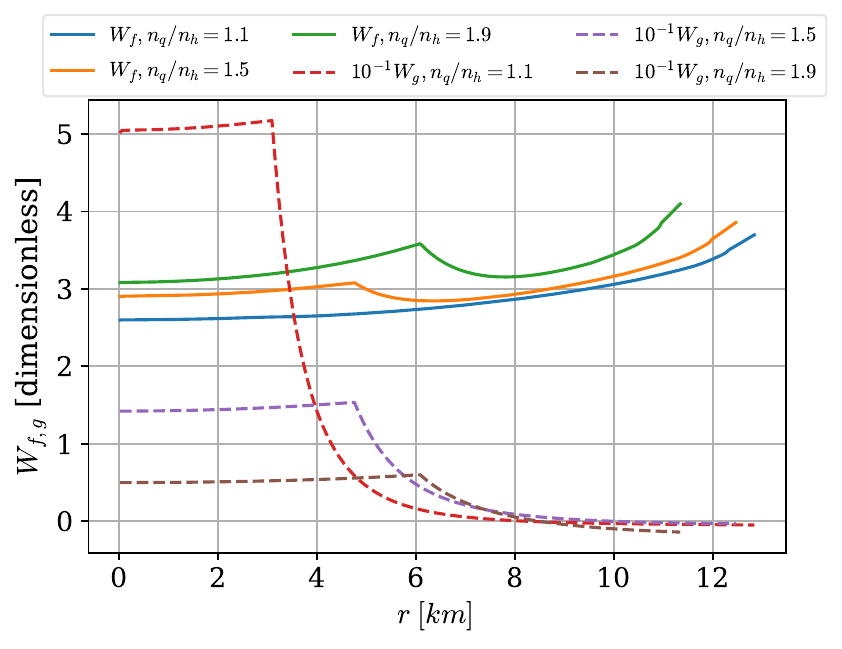}
\caption{Normalized eigenfunctions $W(r)$ for both $f$- and $g$-modes (in geometric units) are shown for hybrid stars with $1.4M_{\odot}$ and quark-hadron density jumps of $n_q/n_h = 1.1$, $1.5$, and $1.9$. The radial component of the $g$-mode, $W_g(r)$, exhibits pronounced variations within the hadronic phase, contrasting with its smoother profile in the quark phase. Interestingly, $W_g(r)$ can become negative near the stellar surface. In contrast, $f$-mode eigenfunctions $W_f(r)$ are much smoother throughout the star and, as expected, present no nodes.
}
\label{W_eigenfrequencies}
\end{figure}

\begin{figure}[!htb]
\includegraphics[width=\columnwidth]{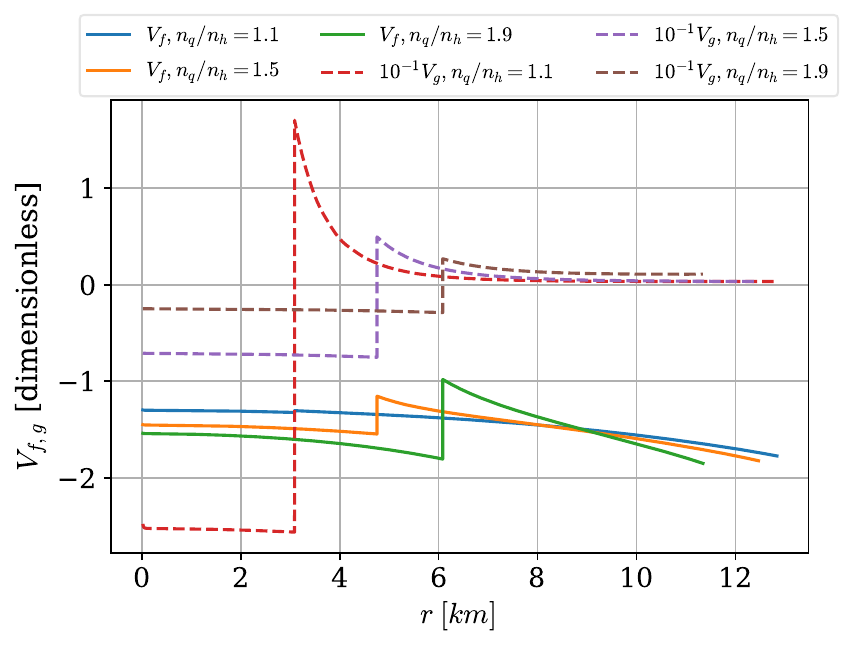}
\caption{Normalized eigenfunctions $V(r)$ for the $f$- and $g$-modes are shown in geometric units for hybrid stars with $1.4M_{\odot}$ and quark-hadron density jumps of $n_q/n_h = 1.1$, $1.5$, and $1.9$. The angular perturbation $V(r)$ exhibits a discontinuity at the quark-hadron interface, stemming from the underlying jump in energy density. The behavior of $V(r)$ differs significantly between the $f$- and $g$-modes: all $V_f(r)$ are negative throughout the star, while $V_g(r)$ remain positive in the hadronic phase. Moreover, the amplitude of the $g$-mode angular perturbations is approximately one order of magnitude larger than that of the $f$-modes.
}
\label{V_eigenfrequencies}
\end{figure}

\begin{figure}[!htb]
\includegraphics[width=\columnwidth]{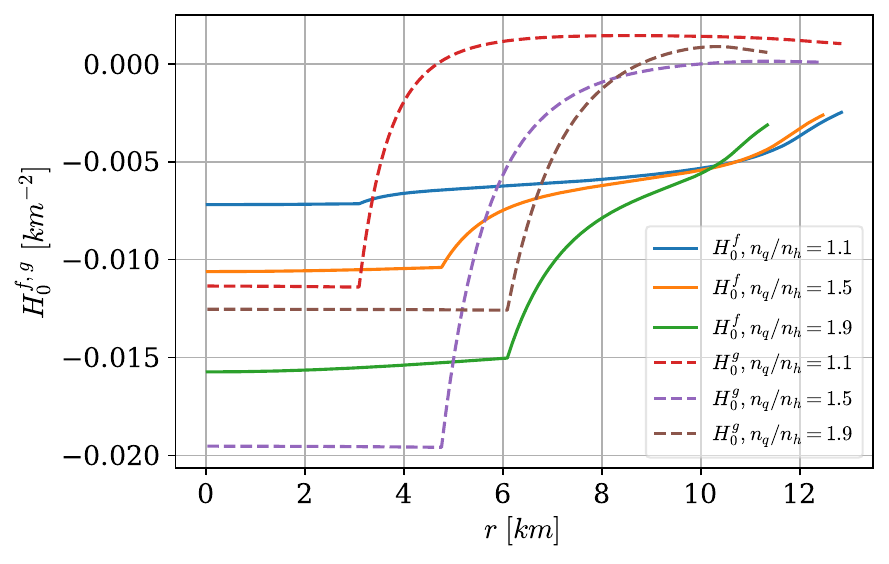}
\caption{The metric perturbations $H_0(r)$ for the $f$- and $g$-modes are shown in geometric units for $1.4\,M_{\odot}$ hybrid stars with $n_q/n_h = 1.1$, $1.5$, and $1.9$. As with the fluid perturbations, the most pronounced variations in $H_0(r)$ occur within the hadronic phase. The $g$-mode solutions for $H_0(r)$ generally exhibit larger amplitudes and can even become positive, in contrast to the typically smaller and negative $f$-mode profiles.}

\label{H0_eigenfrequencies}
\end{figure}

\begin{figure}[!htb]
\includegraphics[width=\columnwidth]{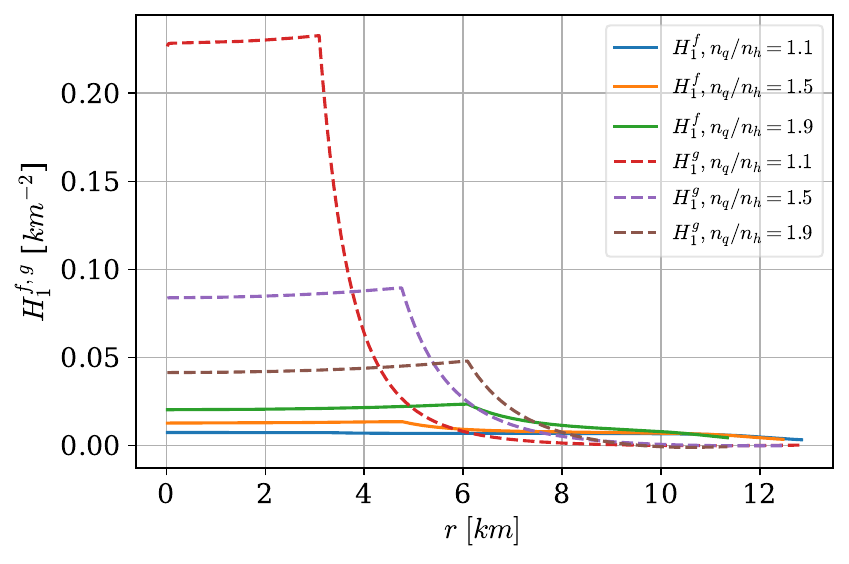}
\caption{The metric perturbations $H_1^f(r)$ and $H_1^g(r)$ are shown in geometric units for $1.4M_{\odot}$ hybrid stars with $n_q/n_h = 1.1$, $1.5$, and $1.9$. As in previous cases, the most significant variations in $H_1(r)$ for both the $f$- and $g$-modes occur within the hadronic phase. The $g$-mode profiles, $H_1^g(r)$, tend to have larger amplitudes compared to their $f$-mode counterparts.}

\label{H1_eigenfrequencies}
\end{figure}

\begin{figure}[!htb]
\includegraphics[width=\columnwidth]{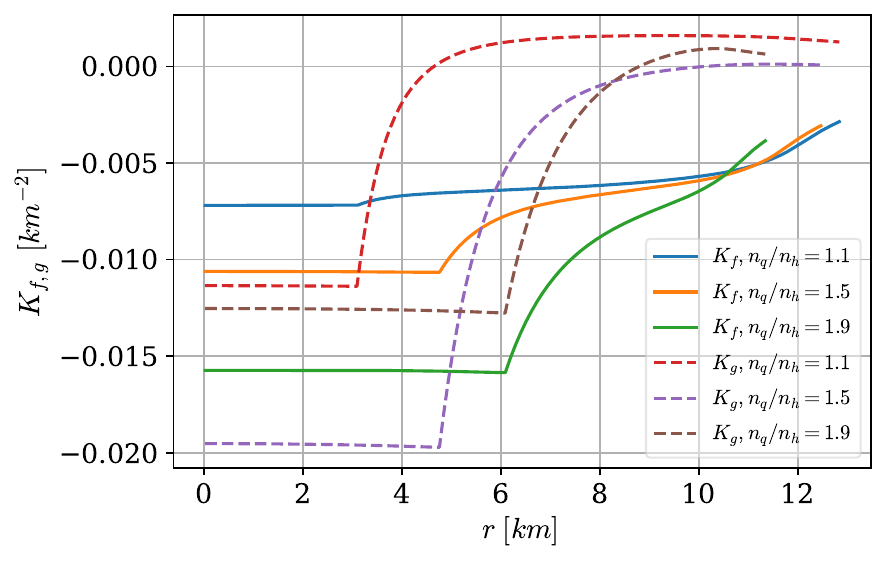}
\caption{The metric perturbations $K(r)$ for both the $f$- and $g$-modes are shown in geometric units for $1.4M_{\odot}$ hybrid stars with $n_q/n_h = 1.1$, $1.5$, and $1.9$. This quantity varies by approximately one order of magnitude less than $H_0(r)$ and $H_1(r)$ for both modes. As with the other metric perturbations, the most significant variations in $K_f(r)$ and $K_g(r)$ occur within the hadronic phase.}
\label{K_eigenfrequencies}
\end{figure}

\begin{figure}[!htb]
\includegraphics[width=\columnwidth]{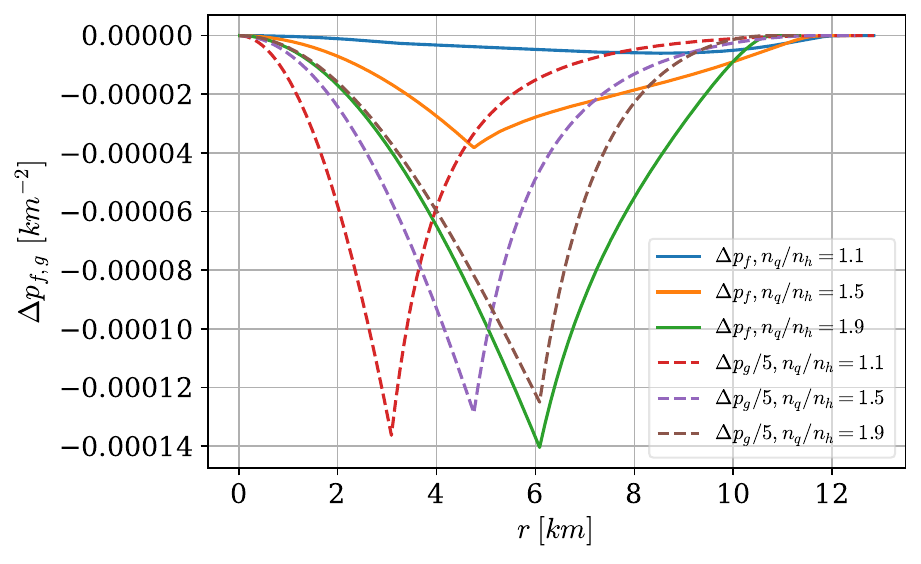}
\caption{The eigenfunctions $\Delta p(r)$ for both $f$- and $g$-modes are shown in geometric units for $1.4M_{\odot}$ hybrid stars with $n_q/n_h = 1.1$, $1.5$, and $1.9$. As expected from the properties of the Lagrangian pressure perturbation, $\Delta p(r)$ vanishes at both the center and the surface of the star for all modes. To satisfy these boundary conditions, the eigenfunctions exhibit nontrivial variations throughout the stellar interior, with more pronounced structure in the $g$-modes. The larger values of $\Delta p(r)$ for the $g$-modes, compared to the $f$-mode, are driven by the action of buoyancy forces at the quark-hadron interface.}
\label{Delta_p_Lag_eigenfrequencies}
\end{figure}

\begin{figure}[!htb]
\includegraphics[width=\columnwidth]{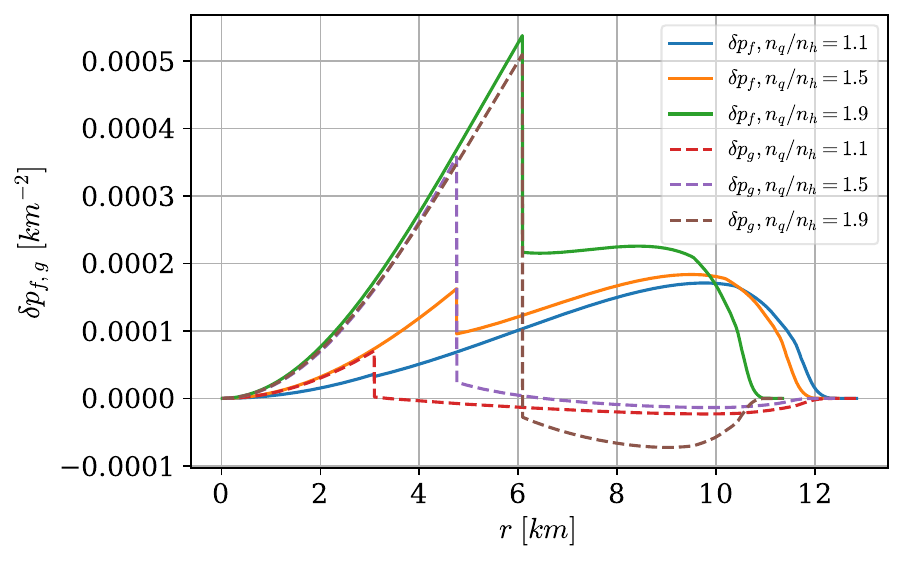}
\caption{The eigenfunctions $\delta p(r)$ for both $f$- and $g$-modes are shown in geometric units for $1.4M_{\odot}$ hybrid stars with $n_q/n_h = 1.1$, $1.5$, and $1.9$. The Eulerian pressure perturbations for both modes exhibit comparable amplitudes within the quark phase for each value of $n_q/n_h$, but differ more noticeably in the hadronic phase. Discontinuities in $\delta p(r)$ are associated with the phase transition. The distinct behavior of the $\delta p$s reflects the different nature of the modes.}
\label{delta_p_Eul_eigenfrequencies}
\end{figure}

To illustrate our strategy, we present solutions for the normalized perturbation variables associated with both $f$- and $g$-modes of $1.4\,M_{\odot}$ hybrid stars, using some selected EOSs shown in Fig.~1 of the main text ($n_q/n_h = 1.1$, $1.5$, and $1.9$). The $g$-modes are more challenging to solve, both in terms of their eigenfrequencies and damping times, whereas $f$-modes are comparatively simpler. For reference, the eigenfrequencies and damping times of the $g$-modes are: for $n_q/n_h = 1.1$, $(586.7~\mathrm{Hz},\, 7.21 \times 10^7~\mathrm{s})$; for $n_q/n_h = 1.5$, $(1287.0~\mathrm{Hz},\, 7.44 \times 10^2~\mathrm{s})$; and for $n_q/n_h = 1.9$, $(1720.2~\mathrm{Hz},\, 16.5~\mathrm{s})$. For the $f$-modes, the corresponding values are: for $n_q/n_h = 1.1$, $(1656.9~\mathrm{Hz},\, 0.26~\mathrm{s})$; for $n_q/n_h = 1.5$, $(1750.3~\mathrm{Hz},\, 0.24~\mathrm{s})$; and for $n_q/n_h = 1.9$, $(2110.9~\mathrm{Hz},\, 0.17~\mathrm{s})$.

As clearly shown in the plots, the perturbation variables vary only slowly within the quark phase for each value of $n_q/n_h$, which extends up to the cusp of the eigenfunction curves and comprises less than approximately $50\%$ of the stellar radius. In contrast, more pronounced variations are observed in the hadronic phase, which begins at the cusp and continues toward the surface. This behavior is expected especially for the $g$-modes, as they are driven by buoyancy restoring forces that are strongest near the quark-hadron interface. Nonetheless, these forces remain active throughout the hadronic phase. More specific details are provided in the caption of each eigenfunction plot (Figs.~\ref{W_eigenfrequencies}--\ref{delta_p_Eul_eigenfrequencies}).

\vskip 1em
\section{Overlap integral for the $g$-modes near the phase transition mass} 
Figure~2 of the main text shows noticeable variations of the overlap integral for small density jumps in the $1.4\,M_{\odot}$ case. These variations are not caused by numerical fluctuations, but rather by the distinct contributions from the quark and hadronic phases to Eq. (2) of the main text, as well as by the nontrivial behavior of the $g$-mode eigenfunctions $V$ and $W$. This is illustrated in Figs.~\ref{W_eigenfrequencies} and \ref{V_eigenfrequencies}, where it is clear that $W$ and $V$ for the $g$-mode can even change sign in the hadronic phase. In addition, they show markedly different amplitudes in the quark and hadronic regions, and are strongly sensitive to the magnitude of the density jump.

\begin{figure}
\includegraphics[width=\columnwidth]{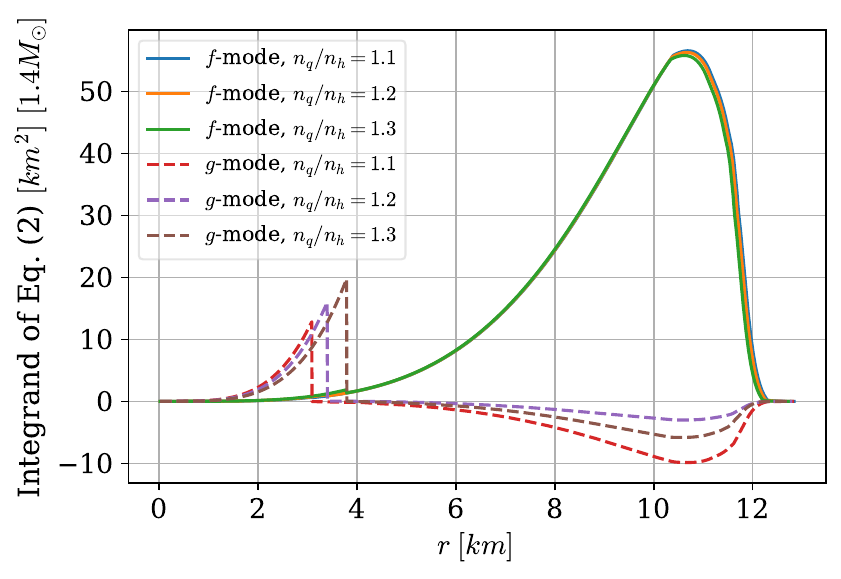}
\caption{Integrand of the overlap integral (Eq.~(2) of the main text) for both the $f$- and $g$-modes in $1.4\,M_{\odot}$ hybrid stars and small density jumps. Note that, for the $g$-modes, the integrand changes sign in the hadronic phase and is very sensitive to the density jump.}
\label{Integrands}
\end{figure}

Masses close to the transition mass—which, in our case, lies very close to $1.4,M_{\odot}$—lead to smaller quark cores, as shown in Fig.~\ref{fig:R_q_R_ratio}. Since the $g$-mode eigenfunctions $V$ and $W$ are more pronounced in the quark phase (see again Figs.~\ref{W_eigenfrequencies} and \ref{V_eigenfrequencies}), a smaller quark core results in a more balanced competition between the quark and hadronic contributions to the overlap integral. Physically, it is well established that the $g$-mode frequency scales with the Brunt–V{\"a}is{\"a}l{\"a} frequency and approaches zero as the density discontinuity decreases \citep{finn1986}. This convergence is nonlinear, and even the presence of a small quark core---located in the innermost, high-density region---can significantly alter the eigenfunction of the $g$-mode. In such cases, the eigenfunction readjusts in response to buoyancy forces throughout the entire star.

This is clearly illustrated in Fig.~\ref{Integrands} for the $1.4,M_{\odot}$ case and small density jumps.
In this regime, the integrand of the overlap integral (Eq.[2] of the main text) for the $g$-mode exhibits both positive and negative contributions, which vary significantly with the density jump. In contrast, for the $f$-mode, the integrand remains nearly unchanged. It is important to note that the integrands are smooth and well-behaved, confirming that these variations are not due to numerical noise. Rather, for small density jumps, the contribution from the quark phase is nearly canceled by that from the hadronic phase.

In contrast, for larger quark cores—associated with larger density jumps and higher masses—the quark phase contribution dominates, yielding a definite sign and a more stable behavior of the overlap integral, as clearly shown in Fig.~2 of the main text.

\begin{figure}
\includegraphics[width=\columnwidth]{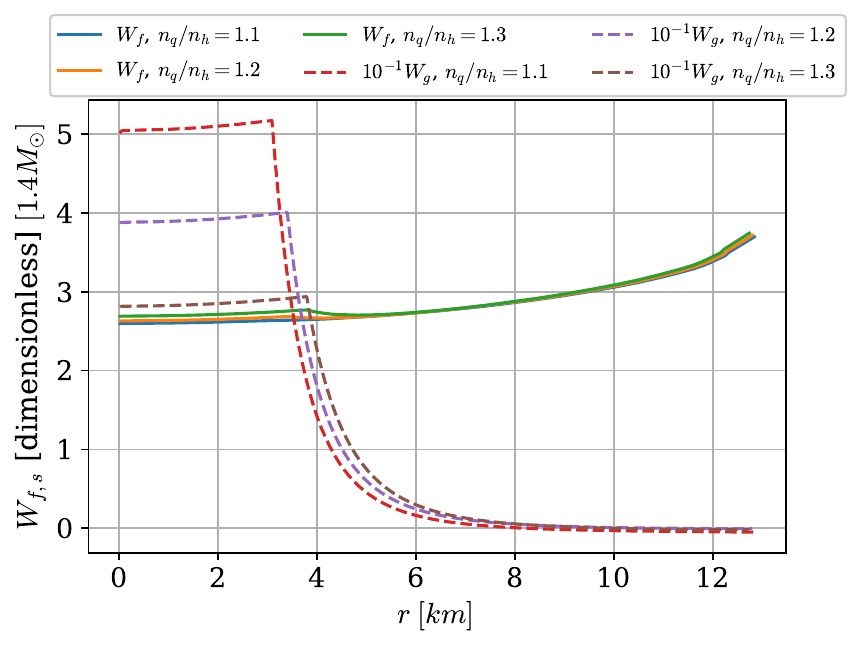}
\caption{$W_g$ and $W_f$ for $n_q/n_h={1.1,1.2,1.3}$ and $M=1.4,M_\odot$. Modest changes in $n_q/n_h$ produce pronounced variations in $W_g$, whereas $W_f$ varies only weakly. Note the sharp decrease of $W_g$ in the hadronic phase, a feature not seen in $W_f$.}
\label{WIntegrands}
\end{figure}

\begin{figure}
\includegraphics[width=\columnwidth]{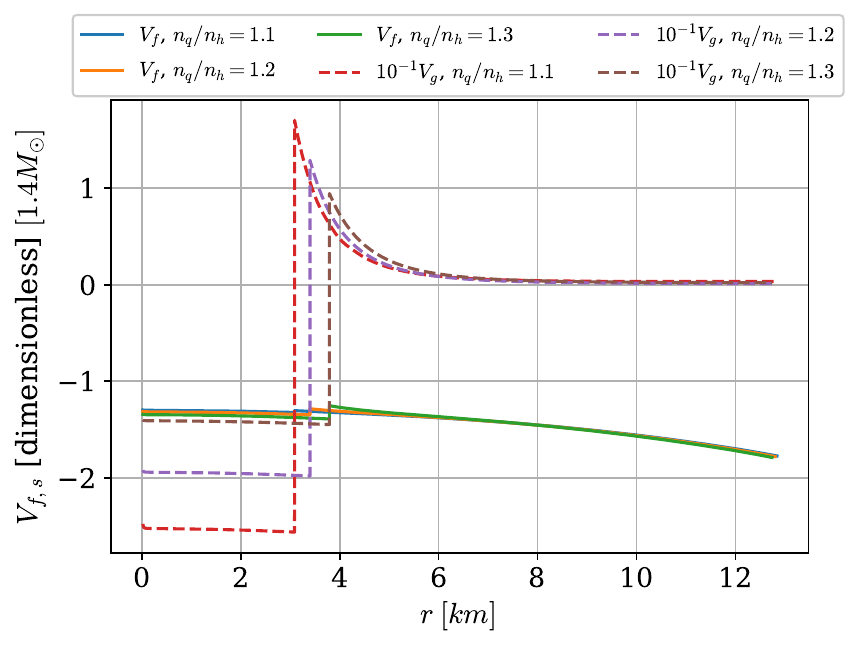}
\caption{$V_g$ and $V_f$ for $n_q/n_h=1.1,,1.2,,1.3$ and $M=1.4\,M_\odot$. Even small changes in $n_q/n_h$ produce sizable variations in $V_g$, in contrast to the very mild response of $V_f$. Analogously to $W_g$, $V_g$ exhibits a sharp decline in the hadronic phase.}
\label{VIntegrands}
\end{figure}

\begin{figure}
\includegraphics[width=\columnwidth]{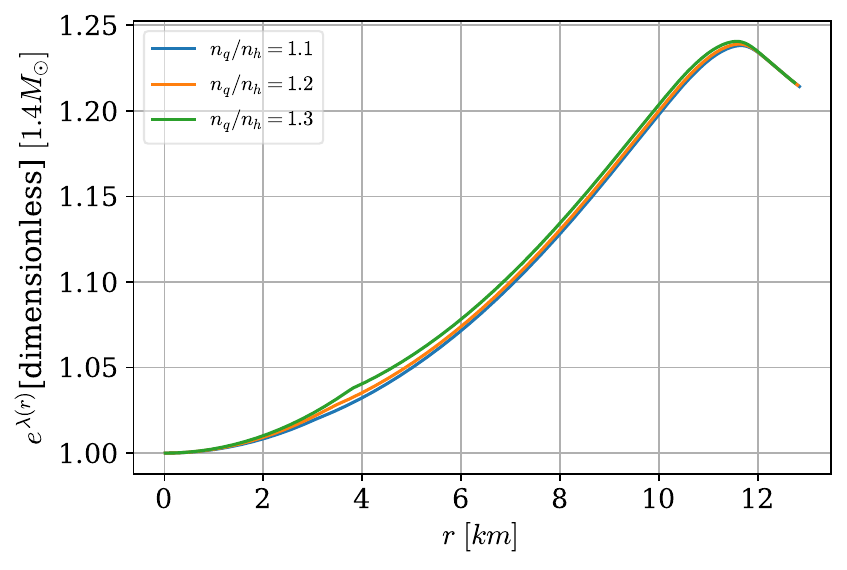}
\caption{Background $rr$-metric component for $n_q/n_h=1.1,1.2,1.3$ and $M=1.4\,M_\odot$. As expected for spacetime geometry, variations are small for modest changes in $n_q/n_h$. Inside the star the relative change in $e^{\lambda (r)}$ can reach up to $\sim25\%$.}
\label{E_lambda_Integrands}
\end{figure}

\begin{figure}
\includegraphics[width=\columnwidth]{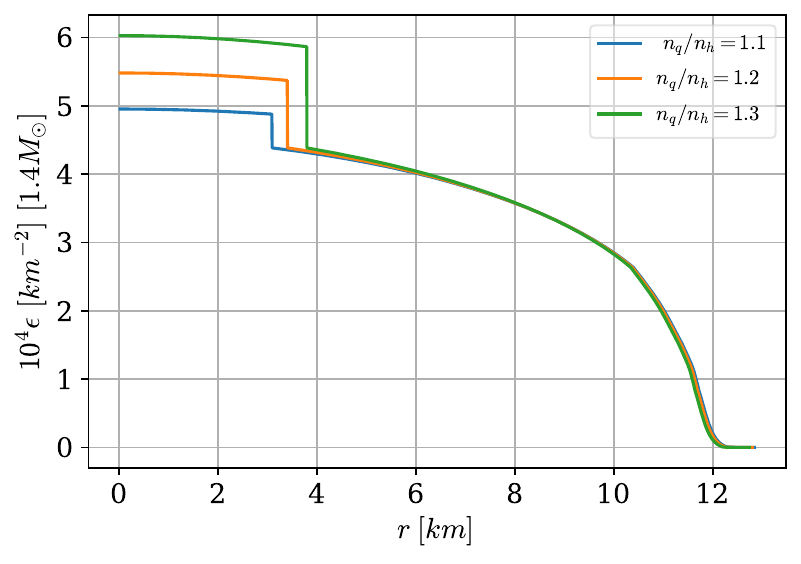}
\caption{Background density profiles for $n_q/n_h=1.1,1.2,1.3$ and $M=1.4\,M_\odot$. The largest changes occur in the quark phase, while they are nearly imperceptible in the hadronic phase.}
\label{rho_Integrands}
\end{figure}

\begin{figure}
\includegraphics[width=\columnwidth]{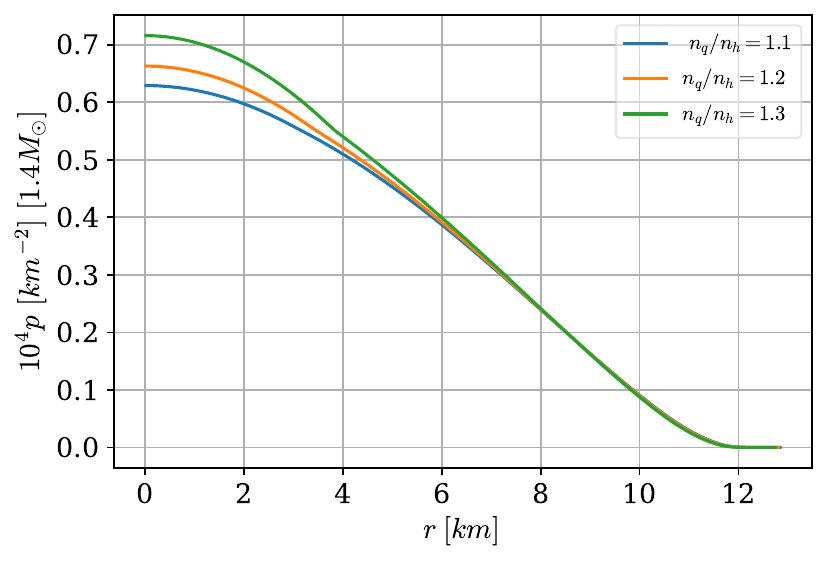}
\caption{Background pressure profiles for $n_q/n_h=1.1,1.2,1.3$ and $M=1.4\,M_\odot$. The largest changes also occur in the quark phase. The differences become imperceptible in the hadronic phase at larger radii than for the density profiles.}
\label{P_Integrands}
\end{figure}

\begin{figure}
\includegraphics[width=\columnwidth]{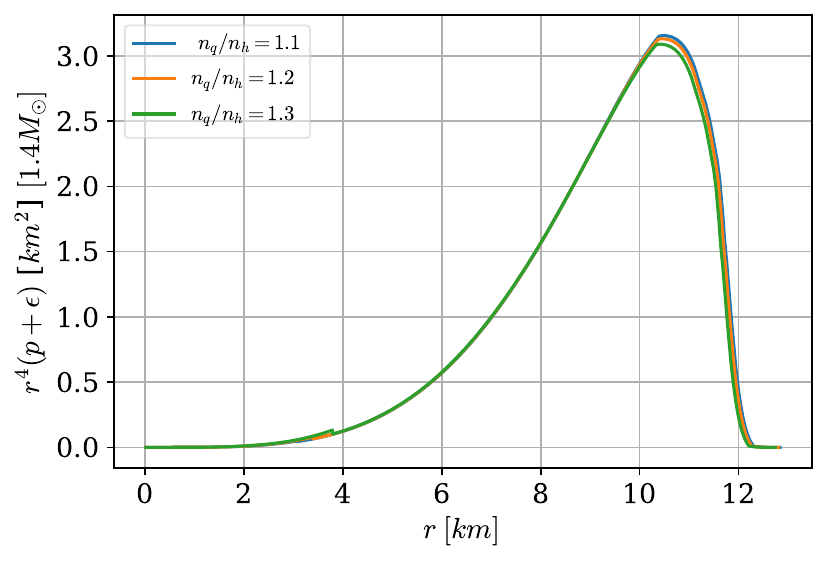}
\caption{Combination $r^4(p+\epsilon)$ for $n_q/n_h=1.1,1.2,1.3$ and $M=1.4\,M_\odot$. Variations of $p$ and $\epsilon$ in the quark phase are largely suppressed because the $r^4$ factor assigns greater weight to the hadronic phase.}
\label{P_r4_Integrands}
\end{figure}

In Figs. \ref{WIntegrands}--\ref{P_r4_Integrands} we decompose the overlap integral into its constituent terms, making their relative weights explicit. As expected, $e^{\lambda}$ varies only weakly for $n_q/n_h=1.1, 1.2,$ and $1.3$. By contrast, $\epsilon$ and $p$ change more markedly within the quark phase. The product $r^4(p+\epsilon)$, however, shows little variation for all density jumps considered, mainly because the factor $r^4$ gives greater weight to the hadronic phase. The terms $W_g$ and $V_g$ vary the most and are therefore mostly responsible for the changes in the overlap integral.

We emphasize that the perturbation problem is fundamentally a boundary condition problem. When a mode's excitation is highly sensitive to a phase transition—and thus to the boundary conditions—it is natural to expect that the $g$-mode eigenfunctions will exhibit non-trivial behavior, especially in the presence of small density jumps near the stellar configuration where the quark core emerges.

\end{document}